\documentclass[12pt]{article}

\usepackage{graphicx}
\usepackage{color}
\usepackage{amsfonts}
\usepackage{amsmath}
\usepackage{natbib}
\usepackage{multirow}
\usepackage{amssymb}

\begin{document}
 
\newcommand{\al}{\mbox{$\alpha$}}
\newcommand{\be}{\mbox{$\beta$}}
\newcommand{\ep}{\mbox{$\epsilon$}}
\newcommand{\gam}{\mbox{$\gamma$}}
\newcommand{\sig}{\mbox{$\sigma$}}

\DeclareRobustCommand{\FIN}{%
  \ifmmode % if math mode, assume display: omit penalty etc.
  \else \leavevmode\unskip\penalty9999 \hbox{}\nobreak\hfill
  \fi
  $\bullet$ \vspace{5mm}}

\newcommand{\calA}{\mbox{${\cal A}$}}
\newcommand{\calB}{\mbox{${\cal B}$}}
\newcommand{\calC}{\mbox{${\cal C}$}}

\newcommand{\muas}{\mbox{$\mu$-a.s.}}
\newcommand{\Nat}{\mbox{$\mathbb{N}$}}
\newcommand{\Rea}{\mbox{$\mathbb{R}$}}
\newcommand{\Prob}{\mbox{$\mathbb{P}$}}

\newcommand{\nin}{\mbox{$n \in \mathbb{N}$}}
\newcommand{\suc}{\mbox{$\{X_{n}\}$}}
\newcommand{\sucP}{\mbox{$\mathbb{P}_{n}\}$}}

\newcommand{\conv}{\rightarrow}
\newcommand{\convn}{\rightarrow_{n\rightarrow \infty}}
\newcommand{\convp}{\rightarrow_{\mbox{c.p.}}}
\newcommand{\convs}{\rightarrow_{\mbox{a.s.}}}
\newcommand{\convw}{\rightarrow_w}
\newcommand{\convd}{\stackrel{\cal D}{\rightarrow}}

\newtheorem {Prop}{Proposition} [section]
 \newtheorem {Lemm}[Prop] {Lemma}
 \newtheorem {Theo}[Prop]{Theorem}
 \newtheorem {Coro}[Prop] {Corollary}
 \newtheorem {Nota}{Remark}[Prop]
 \newtheorem {Ejem}[Prop] {Example}
 \newtheorem {Defi}[Prop]{Definition}
 \newtheorem {Figu}[Prop]{Figure}
 \newtheorem {Tabl}[Prop]{Table}

\title{\sc Models for the assessment of treatment improvement: the ideal and the feasible.}

\author{P. C. \'Alvarez-Esteban$^{1}$, E. del Barrio$^{1}$, J.A. Cuesta-Albertos$^{2}$\\ and C. Matr\'an$^{1}$ \\
$^{1}$\textit{Departamento de Estad\'{\i}stica e Investigaci\'on Operativa and IMUVA,}\\
\textit{Universidad de Valladolid} \\ $^{2}$ \textit{Departamento de
Matem\'{a}ticas, Estad\'{\i}stica y Computaci\'{o}n,}\\
\textit{Universidad de Cantabria}}
%\date{}
\maketitle

\begin{abstract}
Comparisons of different treatments or production processes are the goals of 
a significant fraction of applied research. Unsurprisingly, two-sample problems  
play a main role in Statistics through natural questions such as `Is the the new treatment 
significantly better than the old?'. However, this is only partially answered by some of the 
usual statistical tools for this task. More importantly, often practitioners are  not aware 
of the real meaning behind these statistical procedures. 
We analyze these troubles from the point of view of the order between distributions, 
the stochastic order, showing evidence of the limitations of the usual approaches,
paying special attention to the classical comparison of means
under the normal model. We discuss the unfeasibility of statistically proving stochastic 
dominance, but show that it is possible, instead, to gather statistical evidence to conclude
that slightly relaxed versions of stochastic dominance hold. 
\end{abstract}


\section{Introduction}

Comparison is an essential activity in any field of life, one upon which a significant 
part of human knowledge is founded. In fact, one of the main achievements of mankind --numbers-- 
are just a wonderful sophistication of the comparison process. Whether by curiosity or necessity we are 
continuously involved in comparing objects, leading to assessments like bigger/smaller, 
shorter/taller, better/worse,\dots It is therefore natural the prominent role played in Statistics by 
procedures looking for some kind of ordering. In fact, `Two-Sample Problems',  i.e. comparing two 
populations or two treatments, is probably  the most common situation encountered in statistical 
practice.  Not surprisingly,  every textbook on statistics explores the topic to some extent.

In many cases the practitioner using a two-sample procedure has the goal of
gathering evidence to conclude that a new treatment is better than the old standard.
To fix ideas, let us assume that treatment refers to a particular training program for athletes.
To assess the possible improvement provided by a new training program
the researcher collects some experimental data from athletes training under the two different
programs. Of course not everything comes from the type of training and one expects
that a naturally talented athlete will perform better, whatever the training program,
than a less gifted one. In the simplest case in which performance is measured in terms
of a simple univariate outcome, we can think of a training program as a nondecreasing
transform of the level of natural talent. If, in some scale, this talent level is measured as
$T$ and the different training programs result in $h_{\mbox{old}}(T)$ and $h_{\mbox{new}}(T)$
levels of performance, respectively, we would say that the new treatment is better than the old if
$h_{\mbox{old}}(t) \leq h_{\mbox{new}}(t)$ for all $t$.

What type of conclusion is drawn from the most standard use of the two-sample procedures?
The most commonly used test, namely,  the {\it t}-test, even in the Welch version related to the famous 
Behrens-Fisher problem, would simply aim at rejecting that the mean of $h_{\mbox{old}}(T)$ is greater than
the mean of $h_{\mbox{new}}(T)$. However, as we show in this paper, even under the normality assumptions
implicit in the use of the $t$-test, evidence of a significantly greater mean under the new treatment 
is compatible with a worse performance of, say, 40\% of athletes. We believe that practitioners should be
aware of this fact. We argue in this paper that, most often, they would rather be interested in gathering 
evidence for stochastic dominance than for an increase in mean values.

In this goal of gathering evidence to support the claim that the new treatment yields an improvement over the
old one, we cannot forget that testing hypothesis theory is designed to provide evidence to reject the null hypothesis
and that lack of rejection does not mean evidence for the null. 
While this is a well known fact in the statistical community, in the absence 
of other approaches, practitioners often resort to widely used procedures without full conscience 
of their true meaning. Perhaps the best example of such a situation is the generalized use of goodness of 
fit tests, as the Kolmogorov-Smirnov test, as a way to justify a parametric model assumption, such as normality. 
A  closer example to our present framework is that of  testing homogeneity in a two sample setup. 
In both situations, regardless of the obtained result, we will not be
able to confirm the model but, at best, we would just get lack of statistical evidence to reject it. 
This fact has been pointed out by several authors, notably by \cite{DetteMunk} or \cite{Munk1998}. 
In the particular case of stochastic dominance, we should test the null that stochastic dominance does not 
hold against the alternative that it does hold if we want evidence supporting it (hence, evidence supporting
that the new treatment is better). Unfortunately, no reasonable statistical test can help in this task, see
\cite{Berger} and our discussion in subsection \ref{sovsttest} below.

On the other hand, a model is merely an approximation to reality, so, in order to validate a model, we should be 
conscious of what are the admissible deviations to the model. This is the starting point for the discussion of practical
vs. statistical significance in \cite{Hodges} continued in a series of papers (see e.g. \cite{Rudas},  
\cite{Liu}, \cite{Alv2008}, \cite{Alv2011b}, \cite{Alv2014}) having the common goal of testing the approximate 
validity of statistical hypothesis. In this paper we discuss two relaxations of the stochastic dominance model
for which there are consistent statistical tests. More precisely, we show that there are consistent tests that
allow the practitioner to conclude, up to some small probability of error, that the new treatment is within a small
neighborhood of being better than the old one.

The remaining sections of this paper are organized as follows. In Section 2 we further discuss the convenience of
considering stochastic order rather than growth in mean when trying to assess the improvement given by a new treatment.
Details about the above mentioned lack of valid inferential methods for concluding stochastic dominance are also given,
together with two relaxations of stochastic dominance, which can be used to produce indices of deviation
from the ideal model of stochastic order. We include a subsection that illustrates the behavior of these
indices through the important example of distributions that differ only in changes in location or scale,
showing that growth in the mean is well compatible with a worse performance under a new treatment for a 
very substantial fraction of the population. We invite to a careful inspection of the graphics in figures \ref{normal_restricted} and \ref{normal_order} to get a visual impact of the departures of stochastic dominance measured through such indices, when compared with changes in location and scale in the normal model. Section 3 provides valid inferential methods for gathering statistical evidence that the relaxed stochastic
order models hold, hence, showing that these deviations from the ideal model of stochastic dominance are
tractable models from the point of view of statistical inference. We also provide a simulation study
showing the  performance of the inferential methods in finite samples. Finally, the proofs of some
results introduced in this work are included in an Appendix.

\section{Models of treatment improvement.}

\subsection{Distributional dominance vs. mean comparisons.}\label{sovsttest}

Let us briefly explore the use and real meaning of the most common approach to assess improvement in two 
sample problems. Assuming independence between the samples and normality in the parent distributions, the 
$t$-test is based on the comparison of the means of the distributions. However, relations between the 
means, or any other feature of the distributions, must be cautiously evaluated to assess some kind of 
improvement in a production process or of a treatment with respect to another.

\begin{figure}[htbp] 
\begin{center}
\includegraphics[scale=0.45]{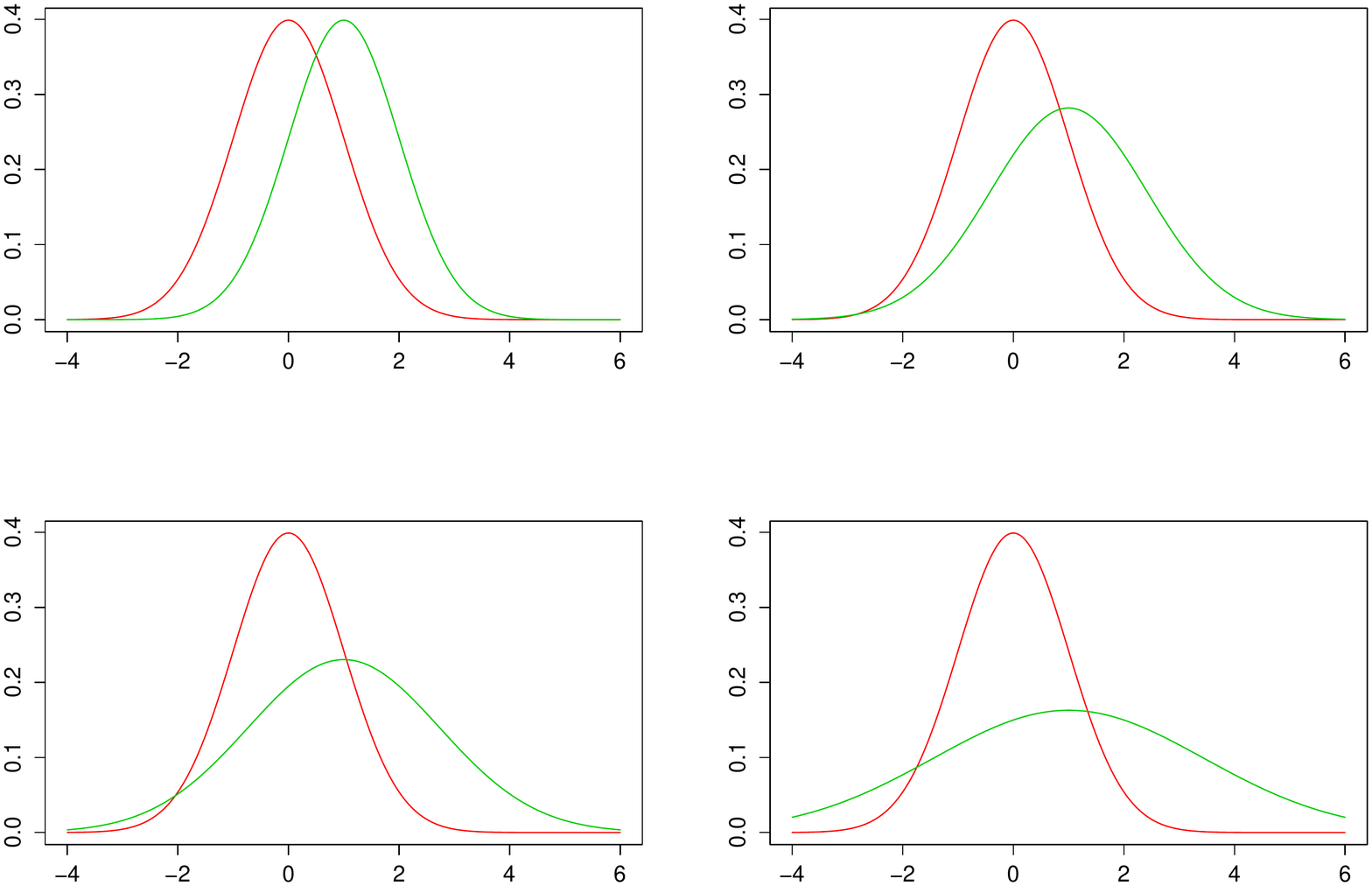} 
\vspace{-10mm}
\caption{Red: standard normal density; green: densities of $N(1,\sigma^2)$, $\sigma^2= 1, 2, 3, 6$.}
\label{densities}
\end{center}
\end{figure}

To motivate our discussion let us assume that the probability laws of the variable of 
interest under two different production processes are normal, say $P_i=N(\mu_i,\sigma_i^2)$, $i=1,2$. 
If we could conclude that $\mu_1 > \mu_2$ we would be only allowed to claim that `in the mean'
the first process produces larger values than the second. To better explain the meaning of such a statement, we 
can resort to the Strong Law of Large Numbers: for large enough samples obtained from both processes, the mean of 
the sample obtained from $P_1$ would be almost surely greater than that of the 
obtained from $P_2$. We should stress the fact that this statement does not depend on the values $\sigma_i$. 
Thus, it is compatible with the situations displayed in Figure \ref{densities}. The relevant 
question is whether these situations are compatible with our intuitive understanding of the statement that 
the first process leads to greater values than the second.

An informal statement such as `men are taller than women' can be better detailed saying that a short 
man would not be as short among women, a medium-sized man would be tall among
women, and a tall man would be even taller among women. These comparisons involve in a natural way 
the relative position or status of every item in both populations. With greater precision, the 
whole comparison involves checking the new status of each element of the first population when
we consider it as an element of the second population. If it is always greater, then we could say 
that the variable in the first population is greater than in the second. 

To analyze if this relation holds, a notable simplification is achieved through a 
previous arrangement of each population ordering their elements  by their status. 
In that way, the relative position of each element in its population is easily obtained, 
and normalization in each population by its size to get comparable status leads to a simple
relation. If $F$ and $G$ are respectively the distribution functions of the variable 
of interest in the first and the second populations, 
and $F^{-1}, G^{-1}$ are the corresponding quantile functions, 
\begin{equation} \label{storderq}  
F^{-1}(t)\leq G^{-1}(t) \mbox{ for every } t\in(0,1),
\end{equation} would mean that the variable in the first population is lower 
than in the second. We recall that for a general distribution function on the real line, 
$F$, the associated quantile function, that we will denote as $F^{-1}$, is defined as
$$F^{-1}(t):=\min \{x \ :  \ t\leq F(x) \} \ \ t \in (0,1).$$

It is well known that the relation (\ref{storderq}) 
is equivalent to the classical  definition of stochastic order, usually attributed to \cite{Lehmann},
but already used at least in \cite{Mann1947}. We say 
that $F$ is stochastically smaller than $G$, and write $F \leq_{st}G$, if 
\begin{equation}\label{storder}
F(x)\geq G(x) \mbox{ for every } x\in\Rea.
\end{equation}
In the econometric literature, where more general classes of stochastic orders 
are considered, usually linked to preferences related with families of utility functions, this 
relation is often invoked as first order stochastic dominance (see, e.g., \cite{Shaked} or \cite{Muller} for several extensions of the concept).

An interesting fact about quantile functions is that if $T$ is uniformly distributed on $(0,1)$
then $F^{-1}(T)$ has distribution function $F$. Let us return for a moment to the discussion in the introduction
and think of $F$ and $G$ as the distribution functions of the performances of athletes training under the old 
and new programs, respectively. Let $T$ be a measure of the natural talent of a randomly  chosen athlete and let $F_T$ be its d.f. 
If we assume that $F_T$ is continuous, then, it is well known that $T^*:=F_T(T)$ is uniform on  $(0,1)$. Therefore, just making a modification on the measurement scale, we have that the variable giving the natural talent  of the athletes is uniform on $(0,1)$. Then, we can see $F^{-1}(T^*)$ and $G^{-1}(T^*)$ as the effects
of the training programs on that natural talent. Hence, $F^{-1}$ plays the role of $h_{\mbox{old}}$ and 
$G^{-1}$ that of $h_{\mbox{new}}$ in the discussion in the introduction and we see from the interpretation there that
the new training program was better than the old if $h_{\mbox{old}}(t)\leq h_{\mbox{new}}(t)$ for all
$t$ coincides with first order stochastic dominance.

\begin{figure}[htbp] 
\begin{center}
\includegraphics[scale=0.45]{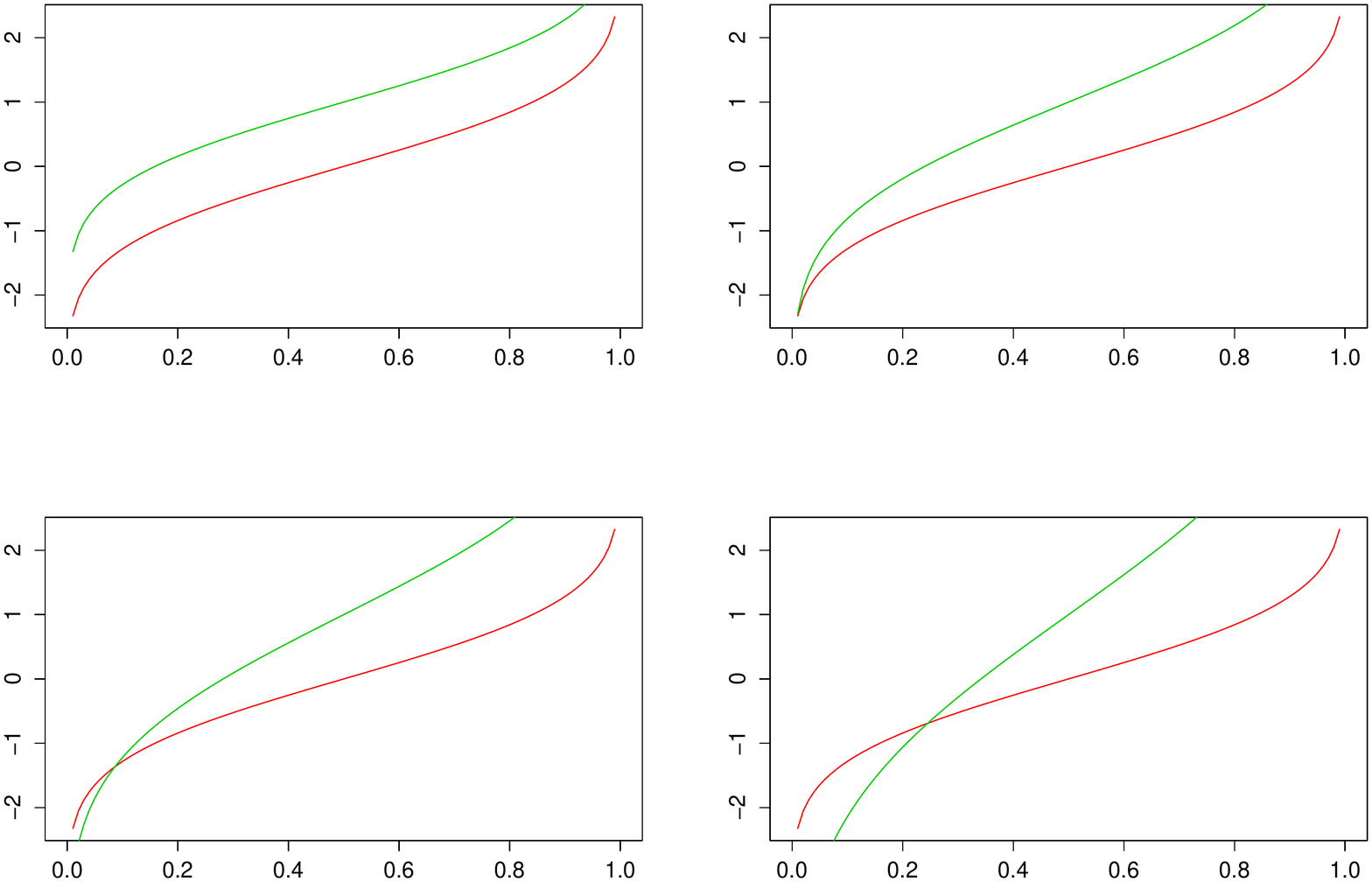} 
\vspace{-10mm}
\caption{Red: standard normal quantile function; green: quantile functions of $N(1,\sigma^2)$, $\sigma^2= 1, 2, 3, 6$.}
\label{quantile_functions}
\end{center}
\end{figure}

In view of the arguments above, we think that a sound answer to question `is the new treatment better than the old'
should be based on the assessment of stochastic order. A look at Figure
\ref{quantile_functions} shows that this cannot be done by simply comparing the means. In fact, the mean is the same for all
distributions in green, but stochastic order only holds in the comparisons $N(0,1)$ vs $N(1,1)$. Specific inferential methods
for assessing $F\leq_{st} G$ are needed.

There is an abundant literature in statistical and econometric journals concerning testing problems related 
to stochastic dominance. Some references, tracing back to \cite{Mann1947}, belong to an
order restricted inferential approach, that is, assuming that stochastic order holds they focus 
on concluding that \textit{strict} stochastic order holds ($F <_{st} G$ if $F(x)\geq G(x)$ for all $x$ with $F(x)>G(x)$ for at least one $x$).
More precisely, they consider the problem of testing the null hypothesis $H_0: F=G$ against the alternative $H_a: F <_{st} G$. 
Obvious as it may be, it is relevant to note that some caution should be adopted in applying these procedures, since,  
both $H_0$ and  $H_a$ can be  simultaneously false.

A different testing problem with a number of references in the literature
(see e.g. \cite{McFadden}, \cite{Anderson}, \cite{Barrett}, \cite{Davidson}, \cite{Linton}, \cite{Linton2}) 
is that of testing the null $H_0$: $F \leq_{st}G$, versus  $H_a$: $F \not \leq_{st}G$.  
This is a kind of goodness-of-fit test. The statistical meaning of not rejecting the null 
is simply to acknowledge that there is not evidence enough to guarantee that stochastic dominance
does not hold. However, this `accepting' the null is sometimes invoked as a guarantee 
that one random variable is stochastically  larger than other, but this is simply wrong.

The available analyses do not address the main goal of gathering statistical evidence 
to assess that stochastic dominance, $F\leq_{st}G$, holds. This would be the natural result of 
rejecting the null in the problem of testing 
\begin{equation}\label{main.test}
H_0: F \not \leq_{st}G, \ \mbox{ versus } H_a: F \leq_{st}G.
\end{equation}
Unfortunately, as already noted in \cite{Berger}, \cite{Davidson2013} and 
\cite{Alv2014}, the statistical assessment of  distributional dominance is impossible because 
small variations in the tails of a distribution could avoid or facilitate a relation of stochastic dominance.
In fact, there is no good $\alpha$-level test for (\ref{main.test}): the `no data' test, rejecting $H_0$ with 
probability $\alpha$ regardless of the data is uniformly most powerful. This is showed in \cite{Berger} in the 
one-sample setting, but the result can be easily generalized to the two-sample setup considered here.
 
\subsection{Relaxations of stochastic dominance.}\label{relaxations}

As it often happens in Statistics, the concept of stochastic dominance is excessively rigid as to try 
to confirm it on the basis of a sample.  It is a too strong assumption in problems in which one 
is inclined to  believe that a population $X$ is somehow smaller than another population $Y$. This 
difficulty seems to be a big justification for the common practice of basing the  comparisons on 
features of the distributions, like the means, when trying to assess some kind of order between 
distributions. For some parametric models, as in the normal model, this approach has the additional 
advantage of leading to true stochastic order for distributions with the same variance. Since optimal 
testing for this problem can be achieved through an exact test, the two-sample $t$-test, the approach 
seems almost perfect. 
For the celebrated Behrens-Fisher problem, when the variances are not assumed equal, 
approximations such  as Welch's proposal give satisfactory enough solutions
to face the comparison of means problem. 
However, even in the normal model, this may be giving a right answer to a wrong question.

Stochastic order is a 0-1 relation. It is either true or false (of course, the same can be said for 
higher order choices of stochastic dominance). 
In the case of normal laws, for instance, stochastic order holds only in the case of equal variances and
increasing means, see Section \ref{location-scale} for related 
results on location-scale models (LS-models in the sequel). However, looking back at
Figure \ref{quantile_functions}, it is tempting to say that the degree of deviation from
stochastic order is higher in the example in the lower right corner than in those in the 
lower left or upper right corners. We could say that in these last cases stochastic dominance 
nearly holds or, even, that `in practice', it holds. Some measurement of the level of agreement
with stochastic order would be helpful.

\begin{figure}[htbp] 
\begin{center}
\includegraphics[scale=.4]{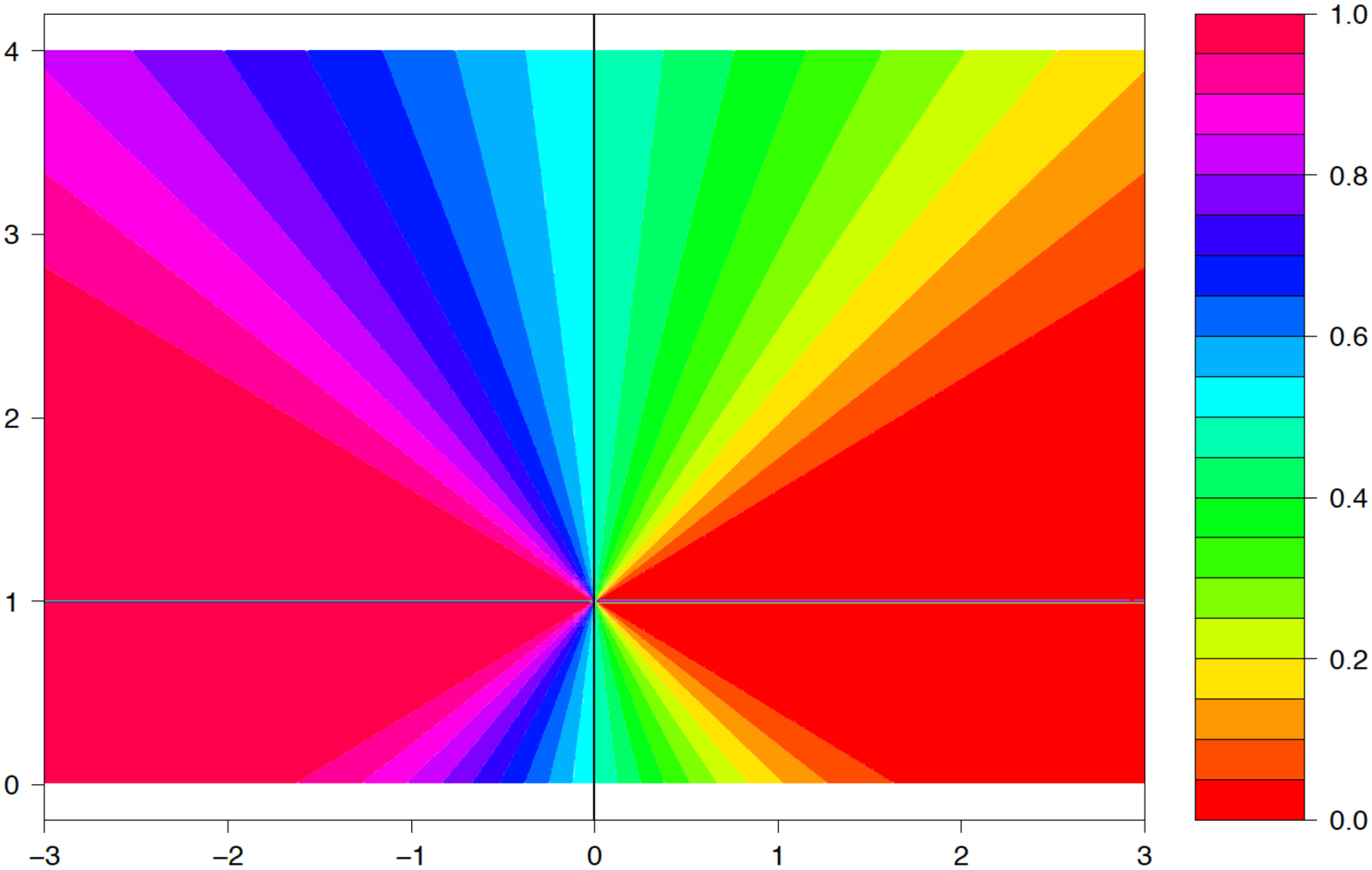} 
\vspace{-10mm}
\caption{Contour-plot of $\gamma(N(0,1),N(\mu,\sigma^2))$ as in (\ref{DefGamma}) for 
different values of $\mu$ (X-axis) and $\sigma$ (Y-axis)} 
 \label{normal_restricted}
\end{center}
\end{figure}

Motivated by the unfeasibility of consistently testing (\ref{main.test}), \cite{Berger} considered 
the idea of `restricted stochastic dominance', which amounts to looking for the relation $F(x)\geq G(x)$ 
on a fixed closed interval, excluding the tails of the sampled distribution. The choice of the interval 
is somewhat arbitrary. The same approach had already been considered in \cite{Lehmann92}, as a weak 
version of the stochastic order, stressing the fact that it is not always appropriate to require that 
the comparison holds for all values of $x$.
An analogous version has been developed in the two-sample setting by \cite{Davidson2013}. 
Given the equivalence between (\ref{storderq}) and (\ref{storder}), it would make sense, as well,
to fix an interval contained in $(0,1)$ and check whether (\ref{storderq}) holds within
this interval. On the other hand, the normalization given by the quantile transform
gives some additional advantage. Rather than facing the arbitrary choice of the interval on which
we want to check that (\ref{storderq}) holds, we can look at the length of the the set
where it does not hold, namely,
\begin{equation}\label{DefGamma}
\gamma(F,G)=\ell\Big( t\in (0,1):\, F^{-1}(t) > G^{-1}(t)\Big),
\end{equation}
where $\ell$ stands for Lebesgue measure. This yields a useful index to measure how far $F$ and 
$G$ are from stochastic order, with $\gamma(F,G)=0$ corresponding to perfect fit. If we turn back to the example of the 
old and new training programs for an athlete, then
$\gamma(F,G)=0.05$ means that 95\% of athletes get better results with the training program associated
to $G$ than with that of $F$. When restricted to a  specific model as the normal 
model (and more generally to LS-models) the computation and the meaning of this index
is easy and very informative. The contour-plot in Figure \ref{normal_restricted} gives a nice 
insight into the fact that moderate and even high levels of disagreement with stochastic order 
(up to $\gamma(F,G)\sim 0.5$) are compatible with an increase in mean, even in the normal model. 
As an example, if $F$ denotes the standard normal law, $N(0,1)$, and $G$ corresponds to the law, $N(.337,1.5^2)$,
then we have $\gamma(F,G)=0.25$. Again, in the training program example, we see that the new program can
yield an improvement in the mean performance of athletes and, yet, result in worse results for 25\%
of them.
We recall that our initial motivation was to discuss the suitability of the usual methods involved in 
the validation of domination or improvement. From Figure \ref{normal_restricted} and the above discussion, 
the inadequacy of the two sample $t$-test to validate a real improvement for most of the population (unless
both distributions satisfy some strong additional assumptions such as normality plus equal variances) 
should be obvious. Of course this is not an objection to the use of, say, the  Welch version of the 
$t$-test for testing an increase in mean. The key point is the meaning of these  mean comparisons 
for the task of showing improvement of treatments or production processes. Later, in Section 
\ref{location-scale} we return to the meaning of definition (\ref{DefGamma}) for normal and, more generally,
LS-models.

While $\gamma(F,G)$ is a natural measure of deviation from stochastic order, it is not the only possible
choice. In \cite{Lesh2002} the authors introduce the so-called Almost Stochastic Dominance which, easily, 
leads to an  index to measure this deviation defining 
\[
\alpha (F,G)= \frac{\int_{\{G>F\}}(G(x) - F(x))dx}{\int_{-\infty}^\infty |F(x) - G(x)|dx}.
\]
 
Although $\gamma(F,G)$ is well defined for any pair of d.f.'s, some assumptions on $F$ and $G$ are needed
in the case of $\alpha(F,G)$. In  \cite{Lesh2002}  the authors require the distributions to be bounded (and limit themselves to the cases 
in which $\alpha (F,G)<.5$). This can be relaxed, but $F$ and $G$ should at least have finite mean. 

This index enjoys nice properties related to the expectation of utility functions (see also \cite{Tsetlin2015}). 
However, it lacks an important property: the stochastic dominance is preserved by monotone functions. This remains true 
for $\gamma (F,G)$ but not for $\alpha(F,G)$. Returning again to the athletes example, let $h$ be 
a strictly increasing function, and assume that we decide to measure the performance of the athletes using the 
values of $h(h_{\mbox{old}})$ and $h(h_{\mbox{new}})$. If $F_h$ and $G_h$ represent the distribution functions of 
the new r.v.'s, then $\gamma(F,G)=\gamma(F_h,G_h)$, while, there is no guarantee that  $\alpha (F,G)=\alpha (F_h,G_h)$. 
We do not pursue further the analysis of the $\alpha$ index in this paper.

Another alternative approach to measure  agreement with stochastic order  
has been introduced in \cite{Alv2015} (see also \cite{Alv2014}).
It is based on  looking for statistical evidence supporting that, 
for a given (small enough)  $\pi,$ there exist mixture decompositions 
\begin{equation}\label{modelocontaminadotwosample1}
\left\{
\begin{matrix}
F & = & (1-\pi)\tilde F+\pi H_F\\
& &\\
G & = & (1-\pi)\tilde G+\pi H_G,
\end{matrix}
\right.
\qquad \mbox{ for some d.f.'s } \tilde F, \tilde G \mbox{ such that } \tilde F\leq_{st} \tilde G.
\end{equation}
We mention some  facts in favor of this approach. First, it allows a robust treatment of the 
problem of stochastic dominance because the decompositions above can be interpreted as contamination 
neighborhoods of some latent distributions $\tilde F$ and $\tilde G$. In this sense we should recall 
that statistical practitioners often process the samples to avoid `rarities' or noise, hence, a methodological 
approach including an adequate treatment for this kind of procedure is helpful. 
On the other hand, by taking $\pi$ large enough, model (\ref{modelocontaminadotwosample1}) always holds 
(the extreme choice $\pi=1$ will always do). The smallest $\pi$ for which a such decomposition is possible
measures
the fraction of the population intrinsically outside the stochastic order model. This provides an index of 
disagreement with the stochastic order model, similar to the lack of fit index introduced in \cite{Rudas} for multinomial models or 
in \cite{Alv2011b} as a relaxation of the homogeneity model. The key fact to use the contamination model
to measure deviation from stochastic order is given by the following result, which is contained in 
\cite{Alv2015}.
\begin{Prop}\label{prop3b}
For arbitrary d.f.'s, $F, G$, and $\pi \in [0,1)$, (\ref{modelocontaminadotwosample1}) holds if and only if
$\pi\geq \pi(F,G)$, where
\begin{equation}\label{pilevel}
\pi(F,G):=\sup_{x\in\mathbb{R}} (G(x)-F(x)).
\end{equation}
\end{Prop}

Notice that, when $\pi(F,G)\in(0,1)$ if $F$ and $G$ have continuous densities $f$ and $g$, respectively, then, there exists $x_0$ such that $\pi(F,G)= G(x_0)-F(x_0)$ and  
$x_0$ satisfies  $f(x_0)=g(x_0)$. Also, as $\gamma(F,G)$, $\pi(F,G)$ is invariant for strictly increasing transformations (see Remark 2.6.1 in \cite{Alv2015}).

A better insight into the meaning of model (\ref{modelocontaminadotwosample1}) is gained through the idea of trimmed
probabilities. An $\alpha$-trimming of a probability, $P$, is any other probability, say $Q$, such that 
$$Q(A)=\int_A \tau dP$$
for some function $\tau$ taking values in $[0,\frac 1 {1-\alpha}]$ and every event $A$. The role of the function $\tau$ is to allow to discard
or downplay the influence of some regions on the sample space (having probability up to $1-\alpha$) on the model, mimicking
the common use in robust statistics of removing disturbing observations. Identifying a probability on the line with its d.f., 
if we write $\mathcal{R}_\alpha(F)$ for the set of trimmings of $F$, then $F =(1-\alpha)\tilde F+\alpha H_F$ for some d.f. $H_F$
if and only if $\tilde F\in \mathcal{R}_\alpha(F)$. Hence, model (\ref{modelocontaminadotwosample1}) holds
if and only if there exist $\tilde F\in \mathcal{R}_\alpha(F)$ and $\tilde G\in \mathcal{R}_\alpha(G)$
such that $\tilde F\leq_{st}\tilde G.$ Even more, this happens if and only if,
after trimming the right tail of $F$  and the left tail of $G$ (removing an $\alpha$ fraction in both cases), 
the resulting $\tilde{F}$ and $\tilde{G}$ satisfy $\tilde F\leq_{st}\tilde G$. We refer to \cite{Alv2015} for details.

\begin{figure}[htbp] 
\begin{center}
\includegraphics[scale=0.26]{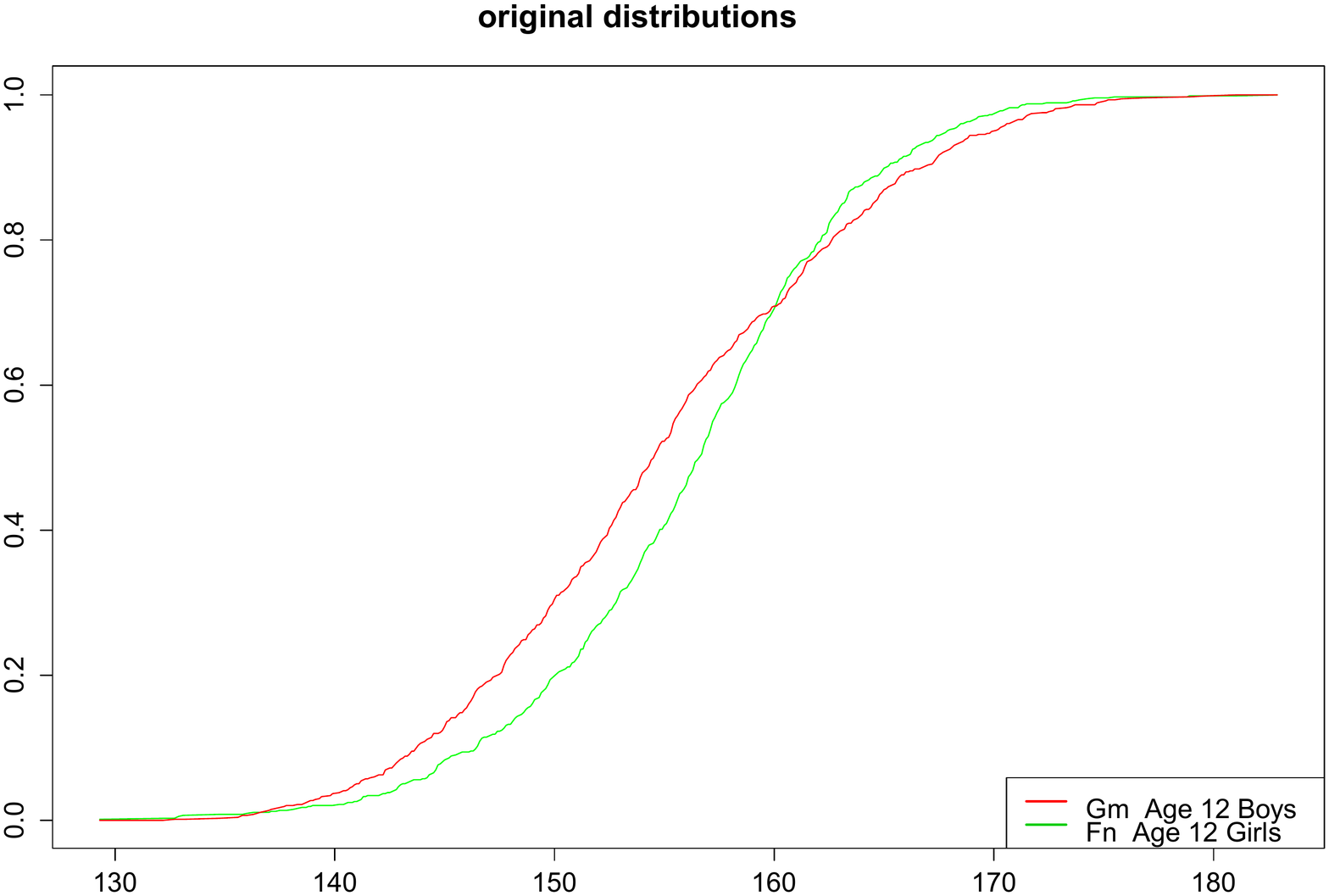} 
\includegraphics[scale=0.26]{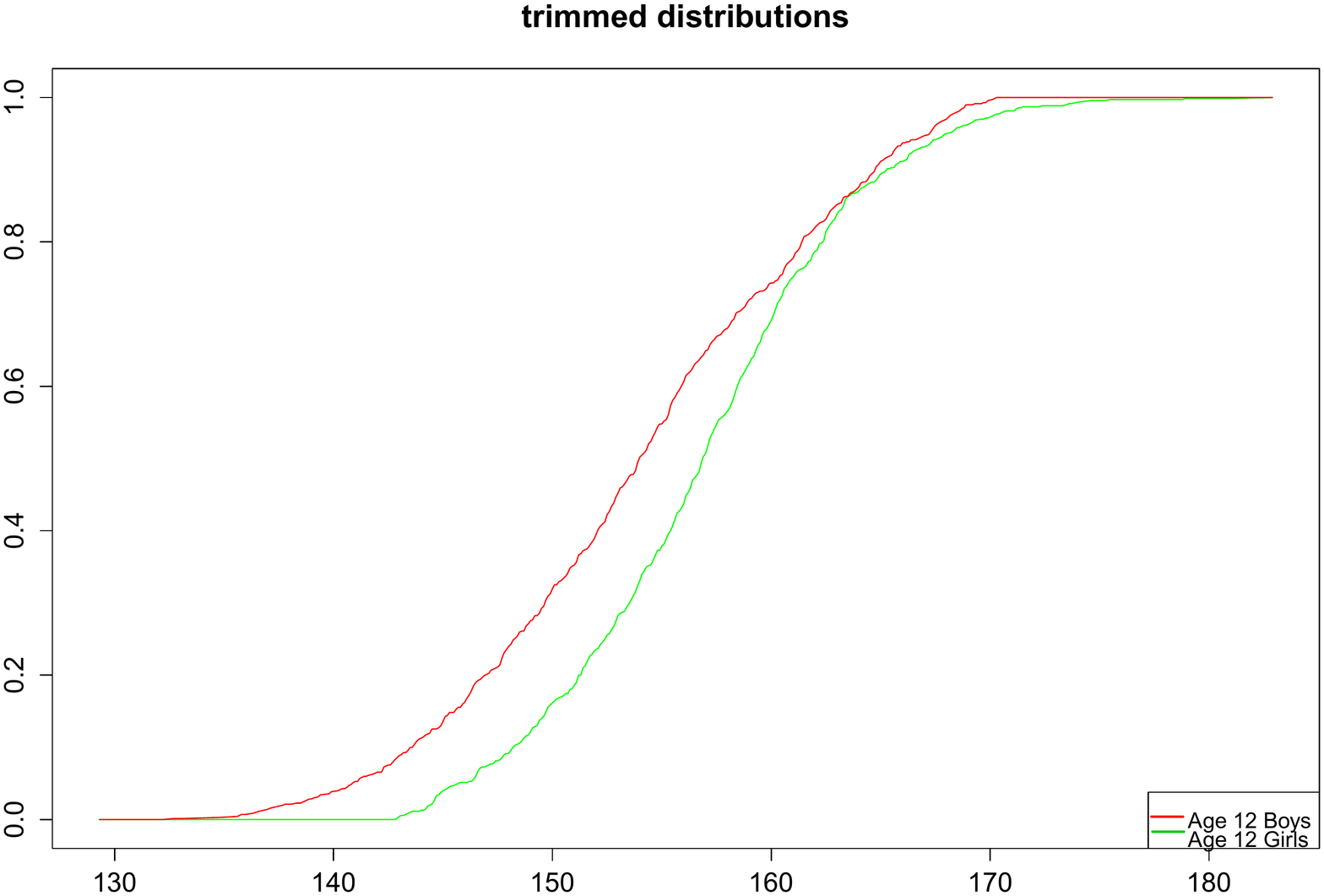} 
\vspace{-5mm}
\caption{Sample d.f.'s of the heights of boys,  $G_m$, and girls, $F_n$, aged 12 in the NHANES dataset.}
\label{antes_despues}
\end{center}
\end{figure}

Figure \ref{antes_despues} shows the comparison of the empirical distributions corresponding to two samples of heights 
of boys and girls (12 years old) from a data set discussed in \cite{Alv2014}. The value $\pi(G_m,F_n)=0.046$, means
that it suffices to trim the fraction $\pi=0.046$ of shortest girls and of taller boys to achieve stochastic order 
between the trimmed distributions and shows, up to $0.046$ contamination, girls (in the sample)
are taller than boys at age 12. On the other hand, to get the stochastic dominance of boys over girls 
we should allow a considerably higher contamination level (because $\pi(F_n,G_m)=0.123$). See \cite{Alv2014} for a more 
detailed analysis of this problem.

\begin{figure}[htbp] 
\begin{center}
\includegraphics[scale=.4]{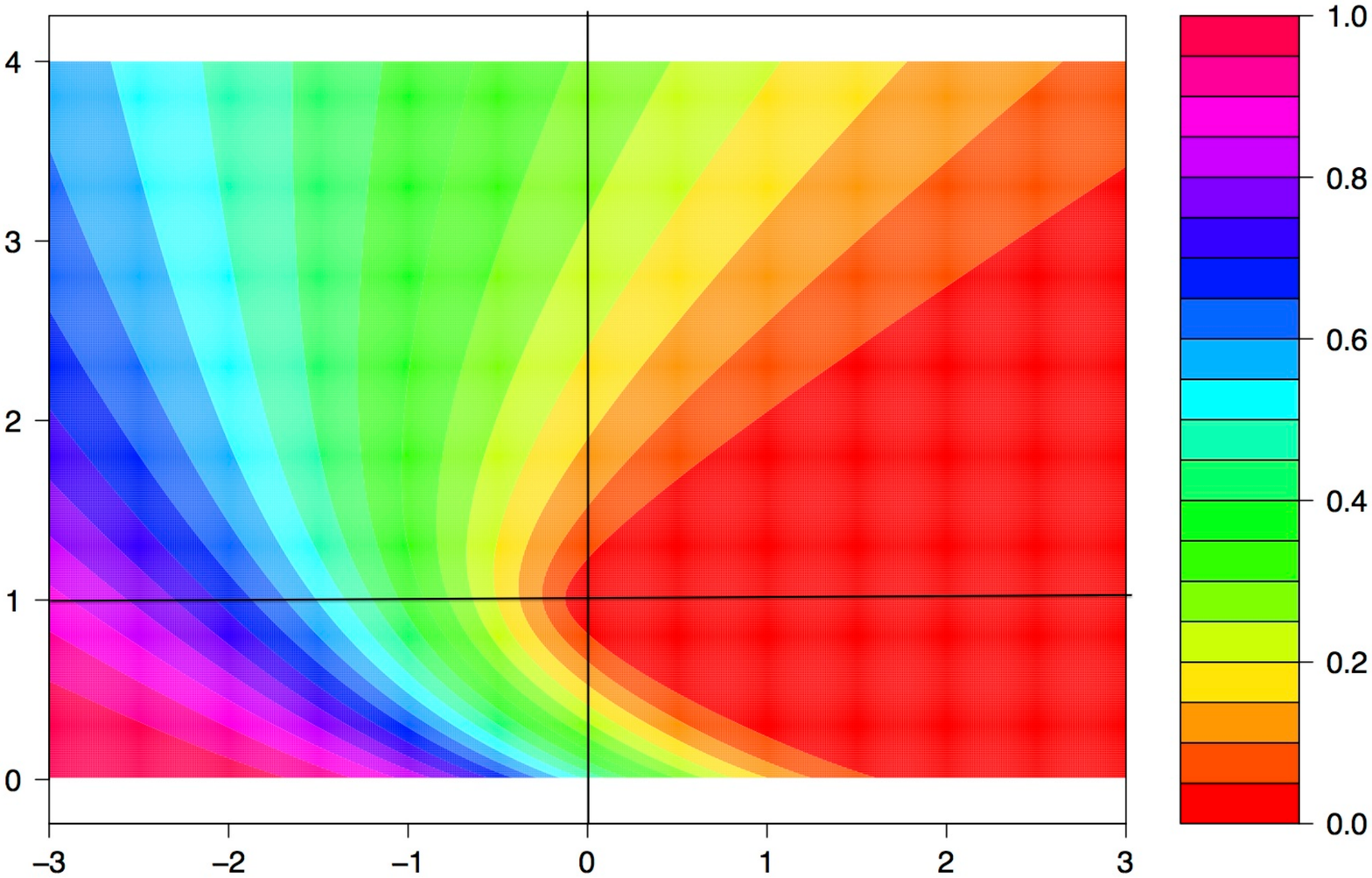} 
\vspace{-10mm}
\caption{Contour-plot of $\pi(N(0,1),N(\mu,\sigma^2))$ as in (\ref{pilevel}) for 
different values of $\mu$ (X-axis) and $\sigma$ (Y-axis)}
 \label{normal_order}
\end{center}
\end{figure}

Coming back to the role of $\pi(F,G)$ as a measure of agreement with the stochastic order, the 
case of the normal model is shown in the contour-plot in Figure \ref{normal_order}. 
As in the case of $\gamma(F,G)$ we see that an increase in the mean is compatible with a high
level of disagreement with respect to stochastic order, that is, with an improvement under the new treatment.
Thus, testing the null assumption that $\pi(F_{N(\mu_1,\sigma_1^2)},F_{N(\mu_2,\sigma_2^2)})\geq 0.05$ against the alternative
$\pi(F_{N(\mu_1,\sigma_1^2)},F_{N(\mu_2,\sigma_2^2)})< 0.05$ would allow to conclude, upon rejection, that
the second treatment results in improvement if we are willing to remove $5\%$ of observations on each side,
while no similar conclusion would be obtained from testing 
$\mu_2\leq\mu_1$ vs $\mu_2>\mu_1$. In Section \ref{Testing approximate stochastic order} 
we will analyze this possibility in the light of the testing procedure developed in \cite{Alv2015}.

A simple comparison between the deviation indices defined in (\ref{DefGamma}) and 
(\ref{pilevel}) arises from the following simple observation. For every couple $(X,Y)$ 
of random variables with marginal d.f.'s $F,G$, we have:
\begin{eqnarray*}\nonumber
G(x)&=&P(Y\leq x, X\leq Y)+P(Y\leq x, X>Y)
 \\
\nonumber 
&\leq& P(X\leq x, X\leq Y)+P(X>Y)  
\\ 
&\leq& 
 F(x)+ P(X>Y), \label{coupling}
\end{eqnarray*}
thus  considering the quantile representations $(X,Y)=(F^{-1},G^{-1})$, since 
$P(X>Y)=\ell(F^{-1}>G^{-1})=\gamma(F,G)$,
from Proposition \ref{prop3b} we get the following statement.

\begin{Prop}\label{pimenorquegamma}
For any pair of d.f.'s, $F$ and $G$,
$$\pi(F,G)\leq \gamma(F,G).$$
\end{Prop}

Now, both $\pi(F,G)$ and $\gamma(F,G)$ are deviation indices from the stochastic order
model, taking values in $[0,1]$ and such that $\gamma(F,G)=0$ if and only if $\pi(F,G)=0$,
with any of these identities being also equivalent to $F\leq_{st} G$. Later, in Section 
\ref{Testing approximate stochastic order} below, we show how to consistently test
$H_0: \pi(F,G)\geq \delta_0$ against $H_a: \pi(F,G)< \delta_0$, which, in case of rejection,
would provide statistical evidence that stochastic order holds aproximately. Under some additional
assumptions we will also provide inferential methods for reaching the more restrictive conclusion
(in view of Proposition \ref{pimenorquegamma}) that $\gamma(F,G)< \delta_0$. 

To conclude this section
we would like to mention that $\pi(F,G)$ and $\gamma(F,G)$ are related to the concepts of trimming and Winsorizing, respectively. Winsorizing and trimming are very popular   robustification procedures in 
Data Analysis,  designed  to avoid an excessive influence of the tails, mainly in presence of 
outliers.  Recall that $\pi(F,G)\leq \delta_0$ if and only if the d.f.'s $\tilde{F}$ and $\tilde{G}$ that we obtain from $F$ and $G$ after removing the $\delta_0$ fraction of the upper tail of $F$ and of the lower tail of $G$, respectively,  satisfy  $\tilde{F}\leq_{st} \tilde{G}$. Winsorizing, in turn, consists in replacing the tails of the distribution with the percentile value from each end.  If we assume that $F^{-1}(t)\leq G^{-1}(t)$ for $t$ in some interval
$(\gamma_1,1-\gamma_2)$ then $\gamma(F,G)\leq \delta_0$ if $\gamma_1+\gamma_2\leq \delta_0$. Of course, 
$F^{-1}\leq G^{-1}$ in $(\gamma_1,1-\gamma_2)$ if and only if the d.f.'s $\tilde{F}$ and $\tilde{G}$
that we obtain from $F$ and $G$ by Winsorizing (both at quantiles $\gamma_1$ and $1-\gamma_2$) satisfy
$\tilde{F}\leq_{st} \tilde{G}$.

\subsection{Stochastic order and location-scale models}\label{location-scale}

In this subsection we will specialize our analysis to the comparisons under a LS-model. In fact, often, and particularly in the simulation study below, we will 
focus on a normal model.

Let $\mathcal{F}_0$ be any d.f. on the real line. A d.f. $F$ is said to belong to the LS-model based 
on $\mathcal{F}_0$ if it satisfies $F(x)=\mathcal{F}_0(\frac{x-\theta}\lambda), \ \ x\in \Rea$ for 
some `location', $\theta \in \Rea$, and `scale', $\lambda>0$ parameters. The dependence on these parameters 
will be included in the notation through the corresponding subindices in the way $F_{\theta,\lambda}$. 
By resorting to the quantile functions, we obtain the characterization
\[%begin{equation}\label{loc_scale}
F_{\theta,\lambda}^{-1}(y)=\lambda \mathcal{F}_0^{-1}(y)+ \theta, \ \ \mbox{ for every } y\in (0,1),
\]%end{equation}
from which the stochastic order $F_{\theta_1,\lambda_1}\leq_{st} F_{\theta_2,\lambda_2}$ would require 
\begin{equation}\label{st_order_loc_scale}
(\lambda_1-\lambda_2)\mathcal{F}_0^{-1}(y) \leq \theta_2-\theta_1 \ \ \mbox{ for every } y\in (0,1).
\end{equation}
Under the assumption that $\mathcal{F}_0$ {\it is continuous and strictly increasing}, (something that we will assume 
from now on in the LS-model setup),
condition (\ref{st_order_loc_scale}) holds if and only if $\lambda_1=\lambda_2$ and $\theta_1 \leq \theta_2$. 
Moreover, two quantile functions in the LS-family have a crossing point, say $y_0$, if and only if 
$\lambda_1 \neq \lambda_2$ and $\mathcal{F}_0^{-1}(y_0)=\frac {\theta_2-\theta_1}{\lambda_1  - \lambda_2}$, therefore, if the crossing point exists, it is unique. In 
other words, the set $\{y: \ F_{\theta_1,\lambda_1}^{-1}(y)\leq F_{\theta_2,\lambda_2}^{-1}(y)\}$ is  
$(0, \mathcal{F}_0(\frac {\theta_2-\theta_1}{\lambda_1-\lambda_2})] $ or $[\mathcal{F}_0(\frac {\theta_2-\theta_1}
{\lambda_1-\lambda_2}),1)$ and $\ell(\{F_{\theta_1,\lambda_1}^{-1} > F_{\theta_2,\lambda_2}^{-1}\})$ is 
$1-\mathcal{F}_0(\frac {\theta_2-\theta_1}{\lambda_1-\lambda_2})$ or $\mathcal{F}_0(
\frac {\theta_2-\theta_1}{\lambda_1-\lambda_2})$ (depending of the sign of $\lambda_1 -\lambda_2$).  
Moreover note that 
\begin{equation}\label{canonical}
\gamma(F_{\theta_1,\lambda_1}, F_{\theta_2,\lambda_2})= 
\gamma(F_{0,1}, F_{{\frac {\theta_2-\theta_1} {\lambda_1}},{\frac {\lambda_2} {\lambda_1}}}),
\end{equation}
hence, we can focus our analysis on comparison to the reference d.f., $\mathcal{F}_0$.

Now, given two d.f.'s $F,G$ in the LS-model, if we are interested in guaranteeing an agreement 
with stochastic dominance of $G$ over $F$ of, say, 95\% for the $\gamma(F,G)$ index, it would suffice 
to consider the crossing point of the quantile functions and check whether the interval corresponding to
$\{F^{-1}\leq G^{-1}\}$ has, at least, length $.95$. For the normal model, when the reference d.f. 
is $\Phi$, the standard normal d.f., a simple computation (which generalizes to any 
LS-model replacing $\Phi$ with the reference d.f., $\mathcal{F}_0$) shows that 
\[%begin{equation}\label{normalRSD}
\gamma(N(0,1),N(\mu,\sigma^2))=1-\Phi\big(\textstyle{\frac{\mu}{|\sigma-1|}} \big),\quad \sigma\ne 1,
\]%end{equation}
while $\gamma(N(0,1),N(\mu,1))=0$ if $\mu\geq 0$ and $\gamma(N(0,1),N(\mu,1))=1$ if $\mu< 0$. This identity
has been used to obtain the contour-plot in Figure \ref{normal_restricted}. We see that 
$\gamma(N(0,1),N(\mu,\sigma^2))$ is constant along rays $\{(\mu,\sigma) : \mu=C|\sigma-1|,\sigma):\, \sigma>0\}$, for some $C>0$, 
and becomes singular at $\mu=0, \sigma=1$. We also note that as $\sigma$ grows bigger $1$ (the case of 
higher variance in the second sample), we can have $\mu>0$ while $\gamma(N(0,1),N(\mu,\sigma^2))\to \frac 1 2$.
This shows again that the conclusion $\mu_{\mbox{new}}>\mu_{\mbox{old}}$ is compatible with a worse
performance with the new treatment for up to 50\% of the population.

We analyze now the behavior of $\pi(F,G)$ under the LS-model. The fact that
\begin{equation}\label{simplif}
\sup_{x \in \mathbb{R}}\left(F_{\theta_2,\lambda_2}(x)-F_{\theta_1,\lambda_1}(x)\right)
=
\sup_{x \in \mathbb{R}}
\left(F_{\frac{\theta_2-\theta_1}{\lambda_1},\frac{\lambda_2}{\lambda_1}}(x)-F_{0,1}(x)\right),
\end{equation}
shows that, as in (\ref{canonical}), we have  
$$
\pi(F_{\theta_1,\lambda_1}, F_{\theta_2,\lambda_2})= 
\pi(F_{0,1}, F_{{\frac {\theta_2-\theta_1} {\lambda_1}},{\frac {\lambda_2} {\lambda_1}}}),
$$
and we can consider only the case $F=F_{0,1}$. There is no simple, general expression for 
$\pi(F_{0,1},F_{\theta,\lambda})$ for every LS-model, since the maximization 
problem in (\ref{simplif}) depends on $\mathcal{F}_0$. In the particular case of the normal model
some elementary computations show that, for $\sigma\neq 1$ and $\mu\geq0$,
$\pi(N(0,1),N(\mu,\sigma^2))=\Phi(\frac{\tilde{x}-\mu}{\sigma})-\Phi(\tilde{x})$,
with
\begin{equation}\label{roots}
\tilde{x}=\frac {\mu \pm \sqrt{\mu^2\sigma^2 + 2(\sigma^2 -1)\log \sigma }}{1-\sigma^2},
\end{equation}
where the positive sign is taken for $\sigma >1$ and the negative for $\sigma<1$.
Also note that for the same values of $\mu$ and $\sigma$,  $\pi(N(\mu,\sigma^2),N(0,1))=\Phi(\tilde{x})-
\Phi(\frac{\tilde{x}-\mu}{\sigma})$, with $\tilde{x}$ the
other solution in (\ref{roots}), a fact that allows also to get the solution for nonpositive $\mu$. 
Of course, when $\sigma=1$ and $\mu\geq 0$, $\pi(N(0,1),N(\mu,1))=0$, while  $\pi(N(\mu,1),N(0,1))$ is 
attained at the only crossing point of both density functions $\tilde{x}=\mu/2$. These computations have been
used to produce the contour-plot in Figure \ref{normal_order}. We see that $\pi(N(0,1),N(\mu,\sigma^2))$ has a
smoother behavior than $\gamma(N(0,1),N(\mu,\sigma^2))$. For a better understanding of the different
roles of $\pi(N(0,1),N(\mu,\sigma^2))$ and $\gamma(N(0,1),N(\mu,\sigma^2))$ we include 
Figure \ref{comparison_orders} below. We see that $\gamma(N(0,1),N(\mu,\sigma^2))$ equals the common
value of the d.f.'s at the crossing point, while $\pi(N(0,1),N(\mu,\sigma^2))$ equals the difference between d.f.'s
at the point where the density functions have a crossing point and the $N(\mu,\sigma^2)$ d.f. is above
the standard normal d.f.. 
In the next section we will use this characterization of $\gamma(F,G)$ in terms of crossing points 
(following a similar approach to that in \cite{Hawkins}) to design a test for the null
$H_0: \, \gamma(F,G)\geq\delta_0$ against the alternative $H_a: \, \gamma(F,G)<\delta_0$.
Similarly, we will discuss how to consistenly test $H_0: \, \pi(F,G)\geq\delta_0$ against the alternative $H_a: \, 
\pi(F,G)<\delta_0$. These will be, according to the discussion above, feasible ways to gather statistical
evidence that for approximate stochastic dominance (that is, to conclude that, essentially, the new treatment is better than the old).
\begin{figure}[htbp] 
\begin{center}
\includegraphics[scale=.4]{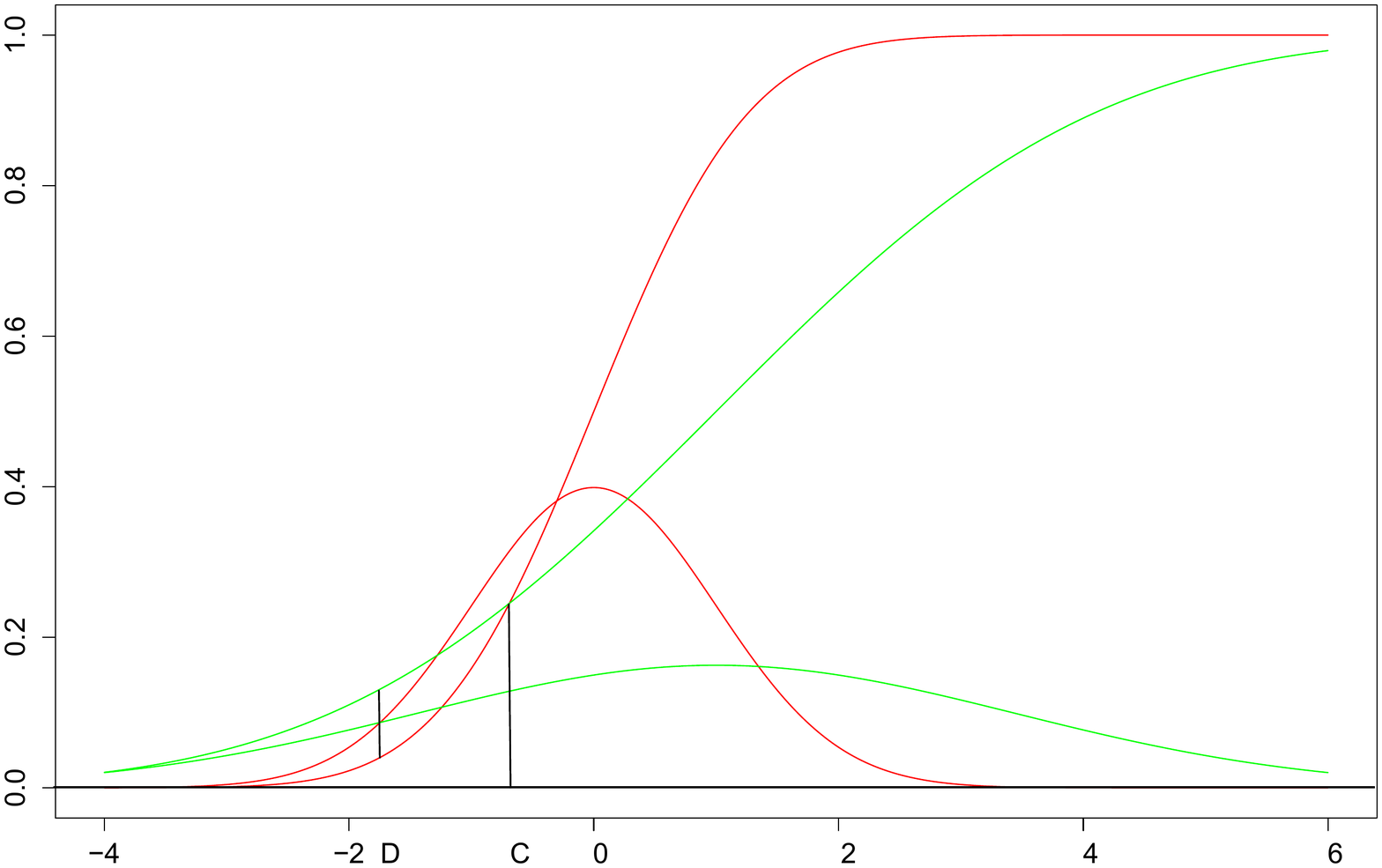} 
\vspace{-10mm}
\caption{Distribution and density functions $N(0,1)$ (red) and $N(1,6)$ (green). Abscissa $C$ (resp. $D$) corresponds to 
crossing points of the distribution (resp. density) functions. 
Lengths of vertical  lines at $C$ and $D$ are $\gamma(N(0,1),N(1,6))$ and $\pi(N(0,1),N(1,6))$, resp.} 
 \label{comparison_orders}
\end{center}
\end{figure}

\section{Testing approximate stochastic order}\label{Testing approximate stochastic order}
In this section we will succinctly analyze feasible test procedures to provide statistical evidence 
of stochastic dominance. We keep in mind that no valid inferential procedure can handle the testing problem
(\ref{main.test}) and, consequently, we have to settle with the less ambitious
testing problems 
\begin{equation}\label{nullmodel2}
H_0:\, \pi(F,G)\geq \pi_0 \mbox{ against the alternative  } H_a:\, \pi(F,G)< \pi_0,
\end{equation}
or
\begin{equation}\label{nullmodel2bis}
H_0:\, \gamma(F,G)\geq \gamma_0 \mbox{ against the alternative  } H_a:\, \gamma(F,G)< \gamma_0,
\end{equation}
with $\pi(F,G)$ and $\gamma(F,G)$ defined in (\ref{pilevel}) and (\ref{DefGamma}), respectively.
We insist that rejection of  the null in (\ref{nullmodel2}) would provide statistical evidence
that stochastic order, up to some small contamination, holds. In (\ref{nullmodel2bis}) rejection would 
lead to guarantee, at the desired level, that treatment $G$ produces better results than treatment
$F$ for at least a fraction of size $1-\gamma_0$ of the population. 

For this we will first include some results, obtained in \cite{Alv2015}, 
that allow to tackle (\ref{nullmodel2}) and present some new results for 
dealing with (\ref{nullmodel2bis}) in a purely 
nonparametric way. Moreover we will include some variations that can be used in the normal model. 
We will finish giving a comparative analysis 
under the normal model, based on a simulation study, providing a picture of the 
feasibility and performance of the different approaches. 
In both (\ref{nullmodel2}) and (\ref{nullmodel2bis}), resorting to  the usual duality between 
one sided testing and confidence bounds, we would 
be interested in obtaining an upper confidence bound, say for $\pi(F,G)$, say for $\gamma(F,G)$. If 
$\hat{U}=\hat{U}(X_1,...,X_n,Y_1,...,Y_m)$  (resp. $\hat{V}$) were an (asymptotic) upper confidence
bound for $\pi(F,G)$ (resp. for $\gamma(F,G)$), rejection of $H_0$ in (\ref{nullmodel2}) (resp. in 
(\ref{nullmodel2bis})) when $\hat{U}< \pi_0$ (resp. when  $\hat{V}< \gamma_0$) would yield a test with
(asymptotically)  controlled type I error probability.
This will be done in two different settings corresponding to nonparametric  and  parametric points 
of view. 

Throughout the section we will assume that
$X_1,...,X_n$ and $Y_1,...,Y_m$ are independent 
i.i.d. r.v.'s  obtained from the d.f.'s $F$ and $G$, respectively. 
We recall that $F_n$ and $G_m$ will denote the sample distribution functions based on the $X'$s 
and $Y'$s samples. As a common assumption in both setups, we will suppose that 
\begin{equation}\label{AssumptionA1}
\mbox{$F$ and $G$ are continuous;\quad $n,m \to \infty$,\quad $\lambda_{n,m}:=\frac n {n+m} \to 
\lambda \in (0,1)$.}
\end{equation}

\subsection{Testing approximate stochastic dominance with the $\pi$ index}\label{subsecpi}

The role of $\pi(F,G)$ in Proposition \ref{prop3b} suggests addressing the testing problem (\ref{nullmodel2}) on the basis of  the  Kolmogorov-Smirnov statistic
$$
\pi(F_n,G_m)=\sup_{x\in\mathbb{R}}(G_m(x)-F_n(x)).$$
Consistency in the strong sense follows from  the Glivenko-Cantelli theorem, which implies
\[%begin{equation}\label{G-C}
\pi(F_n,G_m) \to \pi(F,G) \mbox{ almost surely, as } m, n \to \infty,
\]%end{equation}
while
the asymptotic behavior of its law, under assumption (\ref{AssumptionA1}), was  
obtained by \cite{Ragh73}, extending well known results by Kolmogorov and Smirnov, in the way 
\begin{equation}\label{asympt}
{\textstyle \sqrt \frac {mn}{m+n}} \left(\pi(F_n,G_m)-\pi(F,G)\right) \convw \bar{B}(F,G,\lambda).
\end{equation}
The limit law in (\ref{asympt}) is that of the maximization of a Gaussian Process  on a suitable set
 \[%begin{equation}\label{alternative}
\bar{B}(F,G,\lambda):=\sup_{t \in T(F,G,\pi(F,G))}\left(\sqrt{ \lambda} \ 
B_1(t)- \sqrt{1-\lambda} \ B_2(t-\pi(F,G))\right),
\]%end{equation}
where  $B_1(t)$ and $ B_2(t)$ are i.i.d. Brownian 
Bridges on $[0,1]$, and  the set is \begin{equation}\label{theset}
T(F,G,\pi):= \{t \in [\pi,1]: \ G(x)=t \mbox{ and } F(x)=t-\pi  
\mbox{ for some } x \in \bar{\mathbb{R}} \}.
\end{equation}
Also \cite{Alv2015}  
provides new  results, involving bootstrap and computational issues, as well as 
exponential bounds for both types of error probabilities in testing on the basis of this statistic. Moreover (see the extended version \cite{Alv2014}) it is shown that, for $\alpha \in (0,1/2)$ there are sharp lower  bounds for the $\alpha-$quantiles of the law of $\bar{B}(F,G,\lambda)$, given by the quantiles of a ``least favorable" normal law $N(0,\overline{\sigma}_{\pi(F,G)}^2(\lambda))$ depending of $\lambda$ and of $\pi(F,G)$. 

In fact, (in the 
interesting cases where $\lambda\pi\leq \frac 1 2 \mbox{ and }(1-\lambda)\pi\leq \frac 1 2$) 
taking $\bar{\sigma}_{\pi}^2(\lambda)=
\frac{1}{4}-\pi^2\lambda(1-\lambda),$ rejection of the null in  (\ref{nullmodel2}) when
\begin{equation}\label{proposal1}
\textstyle{\sqrt{\frac{nm}{n+m}}
(\pi(F_n,G_m)-\pi_0) < \bar{\sigma}_{\pi_0}(\lambda_{n,m})\Phi^{-1}(\alpha)},
\end{equation} 
 provides a test of uniform asymptotic level $\alpha$: If $\mathbb{P}_{F,G}$ denotes for the probability when the  samples are obtained from $F$ and $G$, then
 $$\lim_{n\to\infty} \sup_{(F,G)\in H_0} \mathbb{P}_{F,G}\left[  \textstyle{\sqrt{\frac{nm}{n+m}}
(\pi_{n,m}-\pi_0) < \bar{\sigma}_{\pi_0}(\lambda_{n,m})\Phi^{-1}(\alpha)}\right]=\alpha.$$
Moreover, it detects alternatives
with power exponentially close to one (see Proposition 3.3 in \cite{Alv2015}).

Two modifications of (\ref{proposal1}) result in improving the finite sample performance 
of the test. The first relies on the (classical) substitution of the least favorable variance by an appropriate estimation. The second, with a more important effect,  tries to correct the intrinsic bias of $\pi(F_n,G_m)$, a task that often is successfully carried by resorting to the average of a set of boostrap estimates. 
The final proposal, based on these modifications, $\hat{\pi}_{n,m,\mbox{\scriptsize BOOT}}$ of  $\pi(F_n,G_m)$ and $\hat{\sigma}_{n,m}$ of $\bar{\sigma}_{\pi_0}(\lambda_{n,m})$ (see details in \cite{Alv2015}), consists in rejection of 
$H_0:\, \pi(F,G)\geq \pi_0$ if
\begin{equation}\label{proposal3}
{\sqrt{\frac{nm}{n+m}}
(\hat{\pi}_{n,m,\mbox{\scriptsize BOOT}}-\pi_0) < \hat{\sigma}_{n,m}\Phi^{-1}(\alpha)},
\end{equation}
which defines a test of asymptotic level $\alpha$ with quickly
decreasing type I and type II error probabilities away from the null hypothesis boundary. Of course, by defining
 \begin{equation}\label{upperconfidencedirect}
\hat{U}:=\hat{\pi}_{n,m,\mbox{\scriptsize BOOT}}-\textstyle \sqrt{\frac{n+m}{nm}} \hat{\sigma}_{n,m}\Phi^{-1}(\alpha)
\end{equation}
we get an upper bound with asymptotic confidence level at least $1-\alpha$ for $\pi(F,G)$. 

 Let us  notice that  Section 3.4 in \cite{Alv2015} is devoted to the adaptation of these 
statistical tools to the dependent data setup.

\subsection{Testing approximate stochastic dominance with the $\gamma$ index}\label{subsecgamma}

In this subsection we introduce a new procedure for the testing problem (\ref{nullmodel2bis})
under some additional assumption on $F$ and $G$. Basically, we assume that the d.f.'s 
$F$ and $G$ have a single crossing point. This kind of assumption has been considered in some other setups.
\cite{Hawkins} (see also \cite{Chen}) proposed and analyzed 
an estimator of this crossing point, $x^*$. They also stress on the interest of this point for 
comparison of lifetimes under treatments, because, if the r.v. of interest is the survival time, then $x^*$  is the threshold such
that, say, the control subjects have a lower chance of survival to any age $x<x^*$, 
while they have a higher chance of survival to ages $x>x^*$. 
Here we will make the same assumption, but since our interest concerns the common value $\gamma^*=
F(x^*)=G(x^*)$ at this point, we will state it in terms of the quantile functions in the  alternative way: there is a unique $\gamma^*\in (0,1)$
such that $F^{-1}(\gamma^*)=G^{-1}(\gamma^*)$ and, 
\begin{equation}\label{assumption}
F^{-1}(t)-G^{-1}(t)
\mbox{ has opposite signs on } (0,\gamma^*) \mbox{ and on } (\gamma^*,1).
\end{equation} 
In the same spirit as in \cite{Hawkins}, we introduce 
\begin{eqnarray*}\psi(\gamma) &&= \int_0^{\gamma}\left(F^{-1}(t)-G^{-1}(t)\right)dt-
\int_\gamma^1\left(F^{-1}(t)-G^{-1}(t)\right)dt \\ &&= 2\int_0^{\gamma}\left(F^{-1}(t)-G^{-1}(t)\right)dt-(\mu_F-\mu_G),
\end{eqnarray*}
where $\mu_F$, $\mu_G$ denote the means of $F$ and $G$.
The following result gives the link between $\psi$ and  $\gamma(F,G)$.
\begin{Prop}\label{abspsi}
If $F$ and $G$ have finite means and positive density functions on $\mathbb{R}$
and satisfy (\ref{assumption}), then $\gamma^*$ is the unique maximizer of $|\psi(t)|$ on $(0,1)$, 
$\psi(\gamma^*)\ne 0$ and
$$\gamma(F,G)=\left\{ 
\begin{matrix}
\gamma^*&\quad &\mbox{ if } \psi(\gamma^*)>0\\
1-\gamma^*&\quad &\mbox{ if } \psi(\gamma^*)<0.
\end{matrix}
\right. $$
\end{Prop}
This suggests that we consider to estimate $\gamma^*$ by
$$\gamma^*_{n,m}=\min \big(\mbox{argmax}_{\gamma\in (0,1)} |\psi_{n,m}(\gamma)|\big),$$
with $\psi_{n,m}(\gamma)=2 \int_0^{\gamma}\left(F_n^{-1}(t)-G_m^{-1}(t)\right)dt-(\bar{X}_n-\bar{Y}_m)$,
and $\gamma(F,G)$ by
\[%begin{equation}\label{estgamma}
\hat \gamma_{n,m}
=\left\{ 
\begin{matrix}
\gamma_{n,m}^*&\quad &\mbox{ if } \psi_{n,m}(\gamma^*_{n,m})\geq 0\\
1-\gamma_{n,m}^*&\quad &\mbox{ if } \psi_{n,m}(\gamma^*_{n,m})<0.
\end{matrix}
\right. 
\]%end{equation}
The asymptotic behavior of $\hat \gamma_{n,m}$ is given next.
\begin{Prop}\label{gamma.n.m}
Assume that $F$ and $G$ satisfy the conditions of Proposition \ref{abspsi} with densities,
$f$ and $g$ which are continuous in a neighborhood of $x^*$, and such that $f(x^*)\neq g(x^*)$.
Then, if the sample sizes satisfy (\ref{AssumptionA1}),
\[%begin{equation}\label{asympdistgamma}
\textstyle{\sqrt{\frac {nm}{n+m}}}\left(\hat \gamma_{n,m}-\gamma(F,G)\right) \convw N(0,\sigma^2),
\]%end{equation}  
where
\[%begin{equation}\label{varianza}
\sigma^2=\frac{\gamma^*(1-\gamma^*)\left[(1-\lambda)g^2(x^*)+\lambda f^2(x^*)\right]}{\left(g(x^*)-f(x^*)\right)^2}.
\]%end{equation}
\end{Prop}

We can now base our rejection rule for (\ref{nullmodel2bis}) on a bootstrap estimation of $\sigma^2$. 
More precisely, we will reject $H_0:\, \gamma(F,G)\geq \gamma_0$ if
\begin{equation}\label{proposal4}
\textstyle{\sqrt{\frac{nm}{n+m}}
(\hat{\gamma}_{n,m}-\gamma_0) < \hat{\sigma}_{n,m}\Phi^{-1}(\alpha)},
\end{equation}
where $\hat{\sigma}_{n,m}$ is the bootstrap estimator of $\sigma$. 
This rejection rule provides a consistent test of asymptotic level $\alpha$. 
Also, as in (\ref{upperconfidencedirect}), 
\begin{equation}\label{upperconfidencedirectgamma}
\hat{V}:=\hat{\gamma}_{n,m}-\textstyle \sqrt{\frac{n+m}{nm}} \hat{\sigma}_{n,m}\Phi^{-1}(\alpha)
\end{equation}
provides an upper confidence bound for  $\gamma(F,G)$ with asymptotic level $1-\alpha$.
\begin{Nota}We must to point out the singularity of the procedure if the density 
functions coincide at the cross point of the d.f.'s. This fact produces  instability 
of the approach when the densities are very similar. In particular, under a LS-model this will happen when the scales are very similar, a case that should lead to guarantee 
an stochastic dominance on the basis of the estimates of the location parameters.
\end{Nota}

\subsection{The parametric point of view} \label{subsec.param}

If we assume that $F$ and $G$ belong to the same LS-model, with ${\cal F}_0$ continuous and strictly increasing,  then the values $\pi(F,G)$ and $\gamma(F,G)$  can be obtained from the parameters. 
Therefore the considered nonparametric approaches have natural competitors based on the 
estimation of the parameters. Of course, such parametric alternatives will be highly 
nonrobust, specially for small values of the population sizes, that are the interesting 
ones but strongly depend on the tails of the distributions. However we will consider these 
parametric alternatives in order to explore the relative performance of our above proposals 
under perfect conditions. 
In this subsection we assume that the parent distribution functions $F,G$ are 
respectively $F_{N(\mu_1,\sigma_1^2)}, F_{N(\mu_2,\sigma_2^2)},$ although other parametric 
LS-families could be treated in the same way.

By considering the maximum likelihood estimators $\bar X_n, \bar Y_m$ and $S_X^2, S_Y^2$ for 
the involved parameters, we can use the plugin estimators 
\begin{equation}\label{plugin1}
\hat\pi_{n,m}:=\pi(F_{N(\bar X_n,S_X^2)},F_{N(\bar Y_m,S_Y^2)}) \ \mbox{ for } \pi(F,G)
\end{equation}
\begin{equation}\label{plugin2}
\hat\gamma_{n,m}:=\gamma(F_{N(\bar X_n,S_X^2)},F_{N(\bar Y_m,S_Y^2)}) \ \mbox{ for } \gamma(F,G)
\end{equation}
Since these estimators are differentiable functions of the means and the standard deviations of the samples (excepting when $\sigma_1=\sigma_2$, for $\gamma$), 
they will be asymptotically normal with the possible exception of $\gamma$ in the case $\sigma_1=\sigma_2$. 
Avoiding this case, the 
consistency and asymptotic normality would be guaranteed, and once more the bootstrap can be used  
to approximate the distributions of (\ref{plugin1}) and (\ref{plugin2}). The 
derivation of the tests and upper bounds would parallel those in subsections \ref{subsecpi} and 
\ref{subsecgamma}.

\subsection{Some simulations}\label{simulations}
We present now a simulation study that shows the power of the procedures discussed in subsections
\ref{subsecpi} and \ref{subsecgamma} for the assessment of approximate stochastic order. In our simulations 
we have generated pairs of independent i.i.d. random samples of sizes 100, 1000 and 5000 and report the rejection
frequencies of the null hypotheses (\ref{nullmodel2}) and (\ref{nullmodel2bis}) for three choices of $\pi_0$ and 
$\gamma_0$ ($\pi_0,\gamma_0\in \{0.01,0.05,0.1\}$). In all cases we have chosen the underlying distribution functions,
$F$ and $G$, to be normal. This allows to compare the performance of the nonparametric tests (\ref{proposal1}) (with the bootstrap
bias correction) and (\ref{proposal4}) to the parametric procedures discussed in subsection \ref{subsec.param}.
Needless to say, these parametric procedures are inconsistent if the unknown random generators $F$ and $G$ do 
not exactly fit into the parametric model. On the other hand, if $F$ and $G$ satisfy the parametric model then the parametric
procedures should be more efficient. Hence, the performance of these parametric procedures should be taken as an ideal
benchmark to which we compare the performance of the nonparametric, consistent procedures. More extensive
simulations, showing the  performance of the test (\ref{proposal1}) in  different setups, including the least favorable cases, can be found in the Online Appendix 2 to \cite{Alv2015}.

The results are reported in Tables \ref{Tabla.1} and \ref{Tabla.2}. 
In all cases the nominal level of the test is $\alpha=0.05$, $F$ is the $N(0,1)$ distribution and $G$ is a 
$N(\mu,\sigma^2)$. Table \ref{Tabla.1} deals with the testing problem (\ref{nullmodel2}).
Several choices of $\sigma$ ($\sigma \in \{ 0.7, 1, 1.5\}$) have been considered. For each of these three choices
of sigma there are three different values of $\mu$, chosen to make $\pi(F,G)=0.01$ (left column), 
$\pi(F,G)=0.05$ (central column) and $\pi(F,G)=0.1$ (right column). Then, for each combination of sample sizes, of parameters, $\mu$ and $\sigma$,
and of tolerance level, $\pi_0$, there are two reported rejection frequencies, with
the upper row corresponding to the nonparametric test and the lower row to the parametric test.

  \begin{Tabl}\label{Tabla.1}
 Rejection rates for $\pi (N(0,1),N(\mu,\sigma^2)) \geq \pi_0$ at the level $\alpha =.05$ along 1,000 simulations. Upper (resp. lower) rows show the results for nonparametric (resp. parame\-tric) comparisons. The  means for each $\sigma$ have been chosen to satisfy $\pi (N(0,1),N(\mu,\sigma^2)) $ equal to 0.01, 0.05 and 0.10 (respectively: first, second and third columns).
   \end{Tabl}
   \begin{center}
\begin{tabular}{|cr|ccc|ccc|ccc|}
& &\multicolumn{3}{|c}{$\sigma =0.7$}&\multicolumn{3}{|c|}{$\sigma = 1$}&\multicolumn{3}{|c|}{$\sigma = 1.5$}
\\
&Sample&\multicolumn{3}{|c}{means}&\multicolumn{3}{|c}{means}&\multicolumn{3}{|c|}{means}
\\
$\pi_0$&\multicolumn{1}{c|}{size}&$.443$ &$.143$ &$-.050$ &$-.025$ &$-.125$ &$-.251$ &$.770$ &$ .287$ &$-.017$ 
\\
\hline
.01 & 100 & .132 & .010 & .000 & .026 & .011 & .001 & .110 & .009 & .000
 \\
 &  & .095 & .001 & .000 & .002 & .000 & .000 & .113 & .002 & .000
 \\
\cline{2-11}
 & 1000 & .067 & .000 & .000 & .018 & .000 & .000 & .048 & .000 & .000
 \\
 &  & .085 & .000 & .000 & .044 & .001 & .000 & .080 & .000 & .000
 \\
\cline{2-11}
 & 5000 & .044 & .000 & .000 & .016 & .000 & .000 & .049 & .000 & .000
 \\
 &  & .062 & .000 & .000 & .072 & .000 & .000 & .069 & .000 & .000
\\
\hline
.05 & 100 & .379 & .069 & .005 & .071 & .020 & .006 & .431 & .059 & .003
\\
 &  & .675 & .096 & .008 & .230 & .084 & .016 & .730 & .103 & .007
\\
\cline{2-11}
 & 1000 & .993 & .031 & .000 & .397 & .015 & .000 & .997 & .065 & .000
\\
 &  & 1.000 & .051 & .000 & .737 & .060 & .000 & 1.000 & .083 & .000
\\
\cline{2-11}
 & 5000 & 1.000 & .035 & .000 & .979 & .028 & .000 & 1.000 & .037 & .000
\\
 &  & 1.000 & .057 & .000 & 1.000 & .057 & .000 & 1.000 & .052 & .000
\\
\hline
.10 & 100 & .788 & .270 & .033 & .222 & .087 & .027 & .822 & .247 & .046 
\\  
 &  & .960 & .450 & .083 & .543 & .290 & .088 & .978 & .489 & .077 
\\  
\cline{2-11}
 & 1000 & 1.000 & .934 & .035 & .990 & .615 & .028 & 1.000 & .967 & .046 
\\  
 &  & 1.000 & .992 & .059 & 1.000 & .877 & .053 & 1.000 & .996 & .061 
\\  
\cline{2-11}
  & 5000 & 1.000 & 1.000 & .029 & 1.000 & .998 & .032 & 1.000 & 1.000 & .039 
\\  
 &  & 1.000 & 1.000 & .055 & 1.000 & 1.000 & .049 & 1.000 & 1.000 & .053  
\\
\hline
\end{tabular}

\end{center}

We see that the rejection frequencies for the nonparametric procedure show either a reasonable agreement
to the nominal level of the test or are slightly conservative, while the parametric procedure is slightly liberal.
On the other hand, the nonparametric procedure is able to reject the null with remarkably high power. As an illustration, consider, for instance, the block $\sigma=0.7$, $
\pi_0=0.05$.
Within this block the boundary between the null and the alternative hypotheses
corresponds
to the middle column ($\mu=.143$; then $\pi(F,G)=0.05$). The observed
rejection frequencies for
the nonparametric procedure are $.031$ and $.035$ for sample sizes $n=m=1000$
and $n=m=5000$, respectively,
a bit below the nominal level of the test ($\alpha=0.05$). As we move to the
alternative (the left column, $\mu=.443$,
$\pi(F,G)=0.01$) we see that samples of size $n=m=1000$ are enough to reject
the null hypotesis $\pi(F,G)\geq 0.05$
(and conclude that $F$ and $G$ satisfy stochastic order up to less than 5\%
contamination) with high probability
(the observed rejection frequency is $.993$). The worst behaviour in terms of
power corresponds to the case $\sigma=1$ (middle group).
In this case rejection of the null with high power ($90\%$ or higher) requires
sample sizes $n=m=5000$. We note, nevertheless,
that testing for approximate stochastic order is a hard inferential problem
and, on the other hand, sample sizes in this range are not
unusual in many fields of application. As for the parametric procedure
introduced for comparison (bottom rows) we observe that
it presents a better performance in terms of power but it is a bit liberal in some
cases (and recall, again, that it is not a consistent
procedure as we move away from this LS setup).

The results for the testing problem (\ref{nullmodel2bis}) are reported in Table \ref{Tabla.2}.
In this setup the cases $\sigma = \sigma_0$ and $\sigma=1-\sigma_0$ are symmetrical and we have focused on the case 
$\sigma\geq 1$. The case $\sigma =1$ would need a different handling, since $\gamma (N(0,1),N(\mu,1))$ 
only takes the values 0 and 1 depending on when $\mu \geq 0$ or $\mu < 0$. For these reasons we have fixed
$\sigma \in \{1.1, 1.5, 2\}$, choosing then $\mu$ accordingly to get $\gamma(N(0,1),N(\mu,\sigma^2))\in \{.01, .05, 0.10\}$.

We see in this case that the tests (both the nonparametric and parametric) can be  too liberal if the two samples have similar variances.
This is not surprising, since the asymptotic variance in Proposition \ref{gamma.n.m} tends to $\infty$ as $\sigma\to 1$. As we move
away from this singular case we see a  somewhat better degree of agreement to the nominal level of the test, as well as a  generally
good performance in terms of power.

Deviations from the ideal model of stochastic order in terms of the $
\gamma(F,G)$-index admit, arguably, a simpler interpretation than
deviations in $\pi(F,G)$-index, but we see that the assessment of stochastic
order up to a small deviation in $\pi(F,G)$-index is, from the point of view
of statistical inference, a better posed problem, less affected by the
similarity of variances. Finally, we remark that, although both indices are
intrinsically
nonparametric in nature, the scope of $\pi(F,G)$ is considerably larger, since
the single crossing point assumption required for the validity of
the asymptotic theory for the $\gamma(F,G)$-index could be too restrictive for
some real applications.

\newpage

    \begin{Tabl}\label{Tabla.2}
    Rejection rates for $\gamma (N(0,1),N(\mu,\sigma^2)) \geq \gamma_0$ at the level $\alpha =.05$ along 1,000 simulations. Upper (resp. lower) rows show the results for nonparametric (resp. parame\-tric) comparisons. The  means for each $\sigma$ have been chosen to satisfy $\gamma (N(0,1),N(\mu,\sigma^2)) $ equal to 0.01, 0.05 and 0.10 (respectively: first, second and third columns).
   \end{Tabl}

\begin{center}

\begin{tabular}{|cr|ccc|ccc|ccc|}
& &\multicolumn{3}{|c}{$\sigma =1.1$}&\multicolumn{3}{|c|}{$\sigma = 1.5$}&\multicolumn{3}{|c|}{$\sigma = 2$}
\\
&Sample&\multicolumn{3}{|c}{means}&\multicolumn{3}{|c}{means}&\multicolumn{3}{|c|}{means}
\\
$\gamma_0$&\multicolumn{1}{c|}{size}&$.233$ &$.164$ &$.128$ &$1.163$ &$.822$ &$.641$ &$2.326$ &$ 1.645$ &$1.282$ 
\\
\hline
.01 & 100 & .000 & .000 & .000 & .007 & .000 & .000 & .000 & .006 & .000
\\
 &  & .001 & .000 & .000 & .070 & .006 & .000 & .124 & .001 & .000
\\
\cline{2-11}
 & 1000 & .013 & .000 & .000 & .095 & .000 & .000 & .102 & .000 & .000
\\
 &  & .038 & .004 & .000 & .085 & .000 & .000 & .083 & .000 & .000
\\
\cline{2-11}
 & 5000 & .046 & .001 & .001 & .097 & .000 & .000 & .065 & .000 & .000
\\
 &  & .097 & .003 & .000 & .076 & .000 & .000 & .048 & .000 & .000
\\
\hline
.05 & 100 & .015 & .003 & .000 & .338 &  .058 & .015 & .611 &  .089 & .017
\\
 &  & .038 & .008 & .001 & .491 &  .112 & .035 & .764 &  .116 & .016
\\
\cline{2-11}
 & 1000 & .209 & .043 & .007 & .919 &  .088 & .002 & .998 &  .082 & .000
\\
 &  & .319 & .079 & .015 & .992 &  .089 & .001 & 1.000 &  .066 & .000
\\
\cline{2-11}
 & 5000 & .654 & .084 & .008 & 1.000 &  .048 & .000 & 1.000 &  .044 & .000
\\
 &  & .799 & .105 & .011 & 1.000 &  .053 & .000 & 1.000 &  .053 & .000
\\
\hline
.10 & 100 & .061 &  .027 & .007 & .702 &   .256 & .095 & .926 &  .390 & .121
\\
 &  & .090 &  .035 & .016 & .810 &   .310 & .141 & .978 &  .461 & .112
\\
\cline{2-11}
 & 1000 & .540 &  .212 & .073 & 1.000 &   .661 & .084 & 1.000 &  .884 & .066
\\
 &  & .672 &  .258 & .092 & 1.000 &   .795 & .074 & 1.000 &  .988 & .057
\\
\cline{2-11}
 & 5000 & .964 &  .395 & .105 & 1.000 &   .993 & .060 & 1.000 & 1.000 & .049
\\
 &  & .987 &  .437 & .102 & 1.000 &   .999 & .062 & 1.000 & 1.000 & .055
\\
\hline
\end{tabular}

\end{center}

\section*{Appendix: Proofs.}

To the best of our knowledge, the index $\gamma(F,G)$ has been introduced 
just here, thus we include in this appendix some technical details to justify our claims about the asymptotics for the proposed estimator $\hat \gamma_{n,m}$.
\vspace{5mm}

\noindent
\textbf{Proof of Proposition \ref{abspsi}.} We begin noting that $\psi(1)=\mu_F-\mu_G=-\psi(0)$ and also that $\psi$ is differentiable in $(0,1)$,
with derivative $\psi'(t)=2(F^{-1}(t)-G^{-1}(t))$. 

Consider first the case $$F^{-1}(t)>G^{-1}(t), \mbox{ for } t\in (0,\gamma^*), \mbox{ while }
F^{-1}(t)<G^{-1}(t), \mbox{ for } t\in (\gamma^*,1),$$ we have $\gamma(F,G)=\gamma^*$ with 
$\psi'(t)>0$, $t\in (0,\gamma^*)$, while $\psi'(t)<0$, $t\in (\gamma^*,1)$. In particular, $\gamma^*$
is the unique maximizer of $\psi$. 

Now, if $\psi(0)\geq 0$ then $\psi(1)\leq 0$, $\psi(\gamma^*)>\psi(0)\geq 0$ 
and $\psi(t)\in [\psi(1), \psi(\gamma^*)]$ for all $t\in (0,1)$ and we see that $\gamma^*$
is the unique maximizer of $|\psi(t)|$. If, on the other hand, $\psi(0)< 0$ then $\psi(\gamma^*)>\psi(1)>0$, 
$\psi(t)\in [\psi(0), \psi(\gamma^*)]$ for all $t\in (0,1)$ and, again, $\gamma^*$
is the unique maximizer of $|\psi(t)|$. 

In the other case 
$F^{-1}(t)<G^{-1}(t)$, $t\in (0,\gamma^*)$, and
$F^{-1}(t)>G^{-1}(t)$, $t\in (\gamma^*,1)$, we have $\gamma(F,G)=1-\gamma^*$
and, arguing as above, we see that $\gamma^*$ is the unique minimizer of $\psi$ and 
the unique maximizer of $|\psi(t)|$ and satisfies $\psi(\gamma^*)<0$.

\hfill $\Box$

By considering the quantile functions $F_n^{-1}$ associated to the empirical d.f.'s $F_n$ as a random function, we get in a natural way the quantile process defined by
$u_n^F(t)=\sqrt{n}(F_n^{-1}(t)-F^{-1}(t))$ for $t\in (0,1)$. The  study of this statistically meaningful stochastic process was addressed in the second half of the past century. For use in the  proof of Proposition \ref{gamma.n.m} we provide the following lemma.

\begin{Lemm}\label{Lema}
Let $F$ (resp. $G$) be a d.f. with continuous and positive derivative $f$ (resp. $g$) on the interval $[F^{-1}(p)-\varepsilon,F^{-1}(q)+\varepsilon]$ (resp. $[G^{-1}(p)-\varepsilon,G^{-1}(q)+\varepsilon]$). Under the independence assumption on the samples obtained from $F$ and $G$, let $u_n^F(\cdot)$ and $u_m^G(\cdot)$) be the corresponding quantile processes. Then there exist independent versions  $\tilde u_n^F(\cdot)$ and $\tilde u_m^G(\cdot)$, of these processes (with the same joint distribution that the originals)
and independent standard Brownian bridges $\tilde B_1$ and $\tilde B_2$, such that
 $$\sup_{t\in[p,q]} \Big|\Big(\sqrt{\frac{m}{n+m}}\tilde u_n^F(t)- 
\sqrt{\frac{n}{n+m}}\tilde u_m^G(t)\Big)-\Big(\sqrt{1-\lambda} \frac{B_1(t)}{f(F^{-1}(t))}-\sqrt{\lambda}\frac{B_2(t)}{g(G^{-1}(t))}\Big)\Big| \stackrel{a.s.}\to 0.$$ 
\end{Lemm}
Proof: The hypotheses on $F$ and $G$ guarantee (see Example 3.9.24 in \cite{vanderVaartWellner1996}):
$$u_n^F(\cdot) \stackrel{w}\to \frac {B_1(\cdot)}{f(F^{-1}(\cdot))} \mbox{ and } u_m^G(\cdot) \stackrel{w}\to \frac {B_2(\cdot)}{g(G^{-1}(\cdot))}\mbox{ in the space } L^\infty[p,q],$$ where $B_1,B_2$ are independent standard Brownian bridges.
Moreover, the independence of the samples implies that of the quantile processes, thus the joint convergence  $$\Big(u_n^F(\cdot),u_m^G(\cdot)\Big) \stackrel{w}\to 
\Big(\frac {B_1(\cdot)}{f(F^{-1}(\cdot))},\frac {B_2(\cdot)}{g(G^{-1}(\cdot))}\Big).$$
Now we can resort to the Skorohod-Dudley-Wichura almost surely representation theorem (see e.g. Theorem 1.10.3 in \cite{vanderVaartWellner1996}), providing a sequence of pairs $\Big(\tilde u_n^F(\cdot),\tilde u_m^G(\cdot)\Big)\stackrel{d}=\Big(u_n^F(\cdot),u_m^G(\cdot)\Big)$ and a pair $\Big(\frac {\tilde B_1(\cdot)}{f(F^{-1}(\cdot))},\frac {\tilde B_2(\cdot)}{g(G^{-1}(\cdot))}\Big)\stackrel{d}=\Big(\frac {B_1(\cdot)}{f(F^{-1}(\cdot))},\frac {B_2(\cdot)}{g(G^{-1}(\cdot))}\Big)$ such that $\Big(\tilde u_n^F(\cdot),\tilde u_m^G(\cdot)\Big)\stackrel{a.s.}\to\Big(\frac {\tilde B_1(\cdot)}{f(F^{-1}(\cdot))},\frac {\tilde B_2(\cdot)}{g(G^{-1}(\cdot))}\Big)$ in the space $L^\infty[p,q]$. From here, the result is straightforward.

\hfill $\Box$

\medskip
\noindent
\textbf{Proof of Proposition \ref{gamma.n.m}.} We note first that
$$\sup_{\gamma\in[0,1]}|\psi_{n,m}(\gamma)-\psi(\gamma)|\leq \int_0^1 |F_n^{-1}(t)-F^{-1}(t)|dt +
\int_0^1 |G_m^{-1}(t)-G^{-1}(t)|dt\to 0 \mbox{ a.s.}$$
(to check that $\int_0^1 |F_n^{-1}(t)-F^{-1}(t)|dt$ vanishes asymptotically we can use the fact that
it equals the Wasserstein distance between $F_n$ and $F$, see \cite{dBGM1999}). As a consequence
we have that $\gamma_{m,n}^*\to \gamma^*$ a.s.. 

Therefore, in the case $F^{-1}-G^{-1}>0$ in $(0,\gamma^*)$,
$F^{-1}-G^{-1}<0$ in $(\gamma^*,1)$, we will have that a.s. $\psi_{m,n}(\gamma_{m,n}^*)>0$ eventually
and $\hat\gamma_{m,n}=\gamma_{m,n}^*$. Similarly, in the case  $F^{-1}-G^{-1}<0$ in $(0,\gamma^*)$,
$F^{-1}-G^{-1}>0$ in $(\gamma^*,1)$, we will have a.s. that eventually $\hat\gamma_{m,n}=1-\gamma_{m,n}^*$.
Hence, in a probability one set, eventually
$$\textstyle{\sqrt{\frac{n+m}{nm}}}(\hat{\gamma}_{m,n}-\gamma(F,G))=\pm \textstyle{\sqrt{\frac{n+m}{nm}}}(\gamma^*_{m,n}-\gamma^*),$$
with the positive sign in the former case and the negative in the latter. 

By symmetry of the
centered normal laws it suffices to prove the convergence for $\textstyle{\sqrt{\frac{n+m}{nm}}}(\gamma^*_{m,n}-\gamma^*)$.
From this point we assume that we are in the case $F^{-1}-G^{-1}>0$ in $(0,\gamma^*)$,
$F^{-1}-G^{-1}<0$ in $(\gamma^*,1)$ and note that in this case $\gamma^*$ is also the maximizer of $\psi$.

We note also that we can replace $\psi$ by $\phi(\gamma)=\int_{\delta}^\gamma (F^{-1}(t)-G^{-1}(t))dt$ for some fixed
$\delta\in(0,1)$ and still have that $\gamma^*$ is the maximizer of $\phi(\gamma)$, $\gamma\in(\delta,\delta')$ for some other
$\delta'\in(0,1)$. We similarly
set $\phi_{n,m}(\gamma)=\int_{\delta}^\gamma (F_n^{-1}(t)-G_m^{-1}(t))dt$. The assumptions ensure that we can choose
$\delta$ and $\delta'$ such that $f$ and $g$ are continuous and bounded away from 0 in $(\delta,\delta')$.
Then, with the notation introduced for the quantile processes associated to the samples obtained from $F$ and $G$ $$\sqrt{\frac{nm}{n+m}}(\phi_{n,m}(\gamma)-\phi(\gamma))=\int_{\delta}^\gamma \sqrt{\frac{m}{n+m}}u_n^F(t)-
\sqrt{\frac{n}{n+m}}u_m^G(t)dt.$$ 

The application of Lemma \ref{Lema} to the quantile processes
in $L^\infty[\delta,\delta']$,  implies that there are versions of $u_n^F$, 
$u_m^G$, $B_1$ and $B_2$ such that $$\sup_{t\in[\delta,\delta']} \Big|\Big(\sqrt{\frac{m}{n+m}}u_n^F(t)- 
\sqrt{\frac{n}{n+m}}u_m^G(t)\Big)-\Big(\sqrt{1-\lambda} \frac{B_1(t)}{f(F^{-1}(t))}-\sqrt{\lambda}\frac{B_2(t)}{g(G^{-1}(t))}\Big)\Big| \stackrel{a.s.}\to 0.$$  
From this we conclude that for any sequence verifying $\gamma_{n,m}=\gamma^*+o_P(1)$, 
\begin{eqnarray}\nonumber
\lefteqn{\sqrt{\frac{nm}{n+m}} \Big(\phi_{n,m}(\gamma_{n,m})-\phi(\gamma_{m,n})\Big)-\sqrt{\frac{nm}{n+m}} \Big(\phi_{n,m}(\gamma^*)-\phi(\gamma^*)\Big)}\hspace*{4cm}\\
\nonumber
&=& \int_{\gamma^*}^{\gamma_{m,n}} \Big({\textstyle \sqrt{\frac{m}{n+m}}u_n^F(t)-\sqrt{\frac{n}{n+m}}u_m^G(t) }\Big)dt\\
&=& (\gamma_{m,n}-\gamma^*)Z+(\gamma_{m,n}-\gamma^*)o_P(1),\label{fact1}
\end{eqnarray}
with $Z=\sqrt{1-\lambda} B_1(\gamma^*)/f(x^*)-\sqrt{\lambda}B_2(\gamma^*)/g(x^*)$. Since we already obtained the consistency $\gamma_{m,n}^*\to \gamma^*$ a.s., we can apply now Theorem \ref{Wellner} below  to conclude that
$$\sqrt{\frac{nm}{n+m}}(\gamma_{n,m}^*-\gamma^*)=-\frac{Z}{\frac{1}{f(x^*)}-\frac{1}{g(x^*)}}+o_P(1),$$
which completes the proof.

\hfill $\Box$

The following Theorem is a suitable  version of Theorem 3.2.16 in \cite{vanderVaartWellner1996} (see the final comments there leading to this simplified statement). It allows to obtain the asymptotic law of an estimator, like that involved in Proposition \ref{gamma.n.m}, based on an ``argmax" procedure. This kind of argument is one of the best known tools to address the asymptotics of M-estimators (see Section 3.2.4 in \cite{vanderVaartWellner1996})

\begin{Theo}\label{Wellner}(see Theorem 3.2.16 in
\cite{vanderVaartWellner1996})
Let $M_n$ be  stochastic processes indexed by an open interval $\Theta\subset \Rea$ and $M:\Theta \to \Rea$ a deterministic function. Assume that $\theta \to M(\theta)$ is twice continuously derivable at a point of maximum $\theta_0$ with  second-derivative $M''(\theta_0)\neq 0$. Suppose that for some sequence $r_n \to \infty$
\begin{equation*}
r_n\Big(M_n(\tilde \theta_n)-M(\tilde \theta_n)\Big)-r_n\Big(M_n( \theta_0)-M(\theta_0)\Big)\\
=(\tilde \theta_n-\theta_0)Z+o_P\big(|\tilde \theta_n-\theta_0|\big),
\end{equation*}
for every random sequence $\tilde \theta_n=\theta_0+o_P(1)$ and a random variable $Z$. If the sequence $\hat \theta_n \stackrel{P} \to \theta_0$ and satisfies $M_n(\hat{\theta}_n)\geq \sup_{\theta} M_n(\theta)-o_P (r_n^{-2})$ for every $n,$ then
$$r_n(\hat \theta_n-\theta_0)=-\frac Z {M''(\theta_0)}+o_P(1).$$
\end{Theo}

\begin{center}{ \bf ACKNOWLEDGEMENTS}
\end{center}

Research partially supported by the
Spanish Ministerio de Econom\'{\i}a y Competitividad y fondos FEDER, grants  
MTM2014-56235-C2-1-P and MTM2014-56235-C2-2, and  by Consejer\'{\i}a de Educaci\'on de la Junta de Castilla y Le\'on, grant VA212U13.

 \end{document}